\documentstyle[11pt]{amsart}

\newcommand{\nc}{\newcommand}

\nc{\1}{\mbox{\bf 1}}
\nc{\ad}{\operatorname{ad}}
\nc{\Antisym}{\operatorname{Antisym}}
\nc{\Ar}{\operatorname{Ar}}
\nc{\Aut}{\operatorname{Aut}}
\nc{\Coker}{\operatorname{Coker}}
\nc{\dlog}{\operatorname{dlog}}
\nc{\Id}{\operatorname{Id}}
\nc{\Ima}{\operatorname{Im}}
\nc{\Ker}{\operatorname{Ker}}
\nc{\Lie}{\operatorname{Lie}}
\nc{\res}{\operatorname{res}}
\nc{\sgn}{\operatorname{sgn}}
\nc{\Supp}{\operatorname{Supp}}
\nc{\Sym}{\operatorname{Sym}}
\nc{\Ver}{\operatorname{Ver}}

\nc{\barLam}{\bar{\Lambda}}
\nc{\bCC}{\bar{\cal C}}
\nc{\bk}{\mbox{\bf{k}}}
\nc{\bL}{\mbox{\bf{L}}}
\nc{\bomega}{\bar{\omega}}
\nc{\bT}{\mbox{\bf{T}}}
\nc{\bt}{\mbox{\bf{t}}}
\nc{\btt}{\mbox{\bf{\~{t}}}}
\nc{\bu}{\mbox{\bf{u}}}
\nc{\bv}{\mbox{\bf{v}}}
\nc{\bz}{\mbox{\bf{z}}}
\nc{\card}{\operatorname{card}}
\nc{\CA}{\cal A}
\nc{\CB}{\cal B}
\nc{\CC}{\cal C}
\nc{\CI}{\cal I}
\nc{\CJ}{\cal J}
\nc{\CF}{\cal F}
\nc{\CL}{\cal L}
\nc{\CO}{\cal O}
\nc{\CR}{\cal R}
\nc{\CS}{\cal S}
\nc{\CW}{\cal W}
\nc{\dpar}{\partial}
\nc{\fa}{\frak a}
\nc{\fg}{\frak g}
\nc{\fG}{\frak G}
\nc{\fh}{\frak h}
\nc{\fn}{\frak n}
\nc{\hfg}{\hat{\fg}}
\nc{\tB}{\tilde B}
\nc{\tC}{\tilde C}
\nc{\tCI}{\tilde{\CI}}
\nc{\tf}{\tilde f}
\nc{\tfg}{\tilde{\fg}}
\nc{\tfn}{\tilde{\fn}}
\nc{\tI}{\tilde I}
\nc{\tL}{\tilde L}
\nc{\tM}{\tilde M}
\nc{\tomega}{\tilde{\omega}}
\nc{\tOmega}{\tilde{\Omega}}
\nc{\tR}{\tilde R}
\nc{\tit}{\tilde{t}}
\nc{\tx}{\tilde x}
\nc{\vI}{\vec{I}}
\nc{\vJ}{\vec{J}}

\nc{\nen}{\newenvironment}
\nc{\ol}{\overline}
\nc{\ul}{\underline}
\nc{\ra}{\rightarrow}
\nc{\lra}{\longrightarrow}
\nc{\Lra}{\Longrightarrow}
\nc{\Lla}{\Longleftarrow}
\nc{\Llra}{\Longleftrightarrow}
\nc{\hra}{\hookrightarrow}
\nc{\iso}{\overset{\sim}{\longrightarrow}}

\nc{\Thm}[1]{Theorem~\ref{#1}}
\nc{\Prop}[1]{Proposition~\ref{#1}}
\nc{\Lem}[1]{Lemma~\ref{#1}}
\nc{\Cor}[1]{Corollary~\ref{#1}}
\nc{\Conj}[1]{Conjecture~\ref{#1}}
\nc{\Claim}[1]{Claim~\ref{#1}}
\nc{\Defn}[1]{Definition~\ref{#1}}
\nc{\Exa}[1]{Example~\ref{#1}}
\nc{\Rem}[1]{Remark~\ref{#1}}
\nc{\Note}[1]{Note~\ref{#1}}

\nen{thm}[1]{\label{#1}{\bf Theorem.\ } \em}{}
\nen{prop}[1]{\label{#1}{\bf Proposition.\ } \em}{}
\nen{lem}[1]{\label{#1}{\bf Lemma.\ } \em}{}
\nen{cor}[1]{\label{#1}{\bf Corollary.\ } \em}{}
\nen{conj}[1]{\label{#1}{\bf Conjecture.\ } \em}{}
\nen{claim}[1]{\label{#1}{\bf Claim.\ } \em}{}

\nen{defn}[1]{\label{#1}{\bf Definition.\ } }{}
\nen{exa}[1]{\label{#1}{\bf Example.\ } }{}

\nen{rem}[1]{\label{#1}{\em Remark.\ } }{}
\nen{note}[1]{\label{#1}{\em Note.\ } }{}
\nen{exer}[1]{\label{#1}{\em Exercise.\ } }{}

\begin{document}

\title[]{On algebraic equations satisfied by hypergeometric correlators
in WZW models. II.}
\author{Boris Feigin}
\address{B.F.: L.D.Landau Institute for Theoretical Physics, Moscow, Russia}
\author{Vadim Schechtman}
\address{V.Sch.: Dept. of Mathematics, SUNY at Stony Brook, Stony Brook,
NY 11794, USA}
\author{Alexander Varchenko}
\address{A.V.: Dept. of Mathematics, University of North Carolina at
Chapell Hill, Chapell Hill, NC 27599, USA}
\date{April 1994}
\thanks{The second author was supported in part by the NSF grant DMS-9202280.
The third author was supported in part by the NSF grant DMS-9203939.}

\maketitle


\section{Introduction}

\subsection{}  Let $\fg$ be a simple finite dimensional complex Lie algebra;
let $(\ ,\ )$ be an invariant scalar product on $\fg$ normalized in such a way
that $(\theta,\theta)=2$, $\theta$ being the highest root. Fix a positive
integer $k$. Let $L_1,\ldots, L_{n+1}$
be irreducible representations of $\fg$ with highest weights $\Lambda_1,
\ldots, \Lambda_{n+1}$. Suppose that $(\Lambda_i,\theta)\leq k$ for all $i$.

Consider a complex affine $n$-dimensional affine space $\Bbb A^n$ with fixed
coordinates $\bz=(z_1,\ldots,z_n)$.
Consider the space $X_n=\Bbb A^n- \cup_{i,j}\Delta_{ij}$ where
$\Delta_{ij}=\{ (z_1,\ldots,z_n)|z_i=z_j\}$ are diagonals.
According to Conformal field theory, one can define
a remarkable finite dimensional holomorhic vector bundle $\CC(\Lambda_1,\ldots,
\Lambda_{n+1})$ over $X_n$ equipped with a flat connection
(with logarithmic singularities along $\Delta_{ij}$). (We imply that the last
representation "lives" at the point $z_{n+1}=\infty$.)

More precisely, consider a trivial bundle over $X_n$ with a fiber
$(L_1\otimes\ldots\otimes L_{n+1})_{\fg}$. Here we denote by $M_{\fg}$ the
space of coinvariants $M/\fg M$ of a $\fg$-module $M$. Let us denote
this bundle by $\CB (\Lambda_1,\ldots, \Lambda_{n+1})$; it is equipped with
a flat connection given by a system of Knizhnik-Zamolodchikov (KZ)
differential equations, ~\cite{kz}. The bundle
$\CC(\Lambda_1,\ldots, \Lambda_{n+1})$ is a certain quotient of
$\CB (\Lambda_1,\ldots, \Lambda_{n+1})$ stable under KZ connection.

Classically this quotient is described in terms of certain coinvariants
of the tensor product ~{$\bL_1\otimes\ldots\otimes \bL_{n+1}$} where $\bL_i$ is
the irreducible representation of the affine Kac-Moody
algebra $\hfg$ corresponding to $L_i$ and having the central charge $k$
(see  for example \cite{kl} or Sect. \ref{conf} below). The first goal of
the present paper is a precise description of $\CB (\Lambda_1,\ldots,
\Lambda_{n+1})$ in terms of finite dimensional representations $L_i$. More
precisely, the fiber of this bundle at a point $\bz=(z_1,\ldots ,z_{n})$
may be described as follows.

Let $f_{\theta}\in\fg$ be a root vector of weight $-\theta$.
Consider the operator
\begin{equation}
\label{oper}
\bz\cdot f_{\theta}=\sum_{i=1}^{n}z_if_{\theta}^{(i)}:
L_1\otimes\ldots\otimes L_{n}\lra L_1\otimes\ldots\otimes L_{n}
\end{equation}
where $f_{\theta}^{(i)}$ denotes operator acting as $f_{\theta}$ on $i$-th
factor and as identity on the other factors. For a weight $\lambda$ let
$M_{\lambda}$ denote the weight component of a $\fg$-module $M$. The map
{}~(\ref{oper}) induces an operator
\begin{equation}
(\bz\cdot f_{\theta})^{k-(\Lambda_{n+1},\theta)+1}:
(L_1\otimes\ldots\otimes L_{n})_{s_0(\barLam_{n+1})}\lra
(L_1\otimes\ldots\otimes L_{n})_{\barLam_{n+1}}
\end{equation}
where $\barLam_{n+1}$ is the highest weight of the dual module
$L_{n+1}^*$, and $s_0(\barLam_{n+1})=
\barLam_{n+1}+(k-(\Lambda_{n+1},\theta)+1)\theta$. Let us denote this operator
by $\bT(\bz)$.  We prove (see ~\ref{explicit}):

{\bf Theorem.} {\em One has a canonical isomorphism
$$
\CC (\Lambda_1,\ldots,\Lambda_{n+1})_{\bz}\cong \Coker \bT(\bz).
$$}

\subsection{} The second goal of this paper is a construction of a
natural map from $\CC(\Lambda_1,\ldots,\Lambda_{n+1})$ to a certain
bundle of "geometric" origin.

More precisely, let $\Lambda_{n+1}=
\sum_{s=1}^{n}\Lambda_s-\sum_{i=1}^rk_i\alpha_i,\ \alpha_i$ being
simple roots of $\fg$. All $k_i$ are non-negative integers
(otherwise $(L_1\otimes\ldots\otimes L_{n+1})_{\fg}=0$). Set
$N=\sum_{i=1}^rk_i$.
Let us consider the space $X_{n+N}=\Bbb C_{n+N}-\cup_{i,j=1}^{n+N}\Delta_{ij}$;
let us denote coordinates in $X_{n+N}$ by $z_1,\ldots,z_n,t_1,\ldots,t_N$.
We have a projection to the first coordinates $p_N: X_{n+N}\lra X_n$.

Following ~\cite{sv}, define the flat connection on the trivial one-dimensional
vector bundle over $X_{n+N}$ by the $1$-form
$$
\omega=\sum_{s>s'}\frac{(\Lambda_s,\Lambda_{s'})}{k+g}\dlog(z_s-z_{s'})+
\sum_{s,i}-\frac{\alpha_{\pi(i)},\Lambda_s}{k+g}\dlog(t_i-z_s)+
\sum_{i>j}\frac{(\alpha_{\pi(i)},\alpha_{\pi(j)})}{k+g}\dlog(t_i-t_j)
$$
where $g$ is the dual Coxeter number of $\fg$, $\pi: \{ 1,\ldots,N\}
\lra \{ 1,\ldots, r\}$ is any map with $\card (\pi^{-1}(i))=k_i$ for all $i$.
Let us denote the trivial bundle equipped with this connection by
$\CL=\CL(\Lambda_1,\ldots,\Lambda_{n+1})$.
The product of symmetric groups
$\Sigma=\Sigma_{k_1}\times\ldots\times\Sigma_{k_r}$
acts naturally fiberwise on the pair $(X_{N+n},\CL)$.

For each $\bz\in X_n$ consider the De Rham cohomology
$H^N(p_N^{-1}(\bz),\CL_{p_N^{-1}(\bz)})$; these spaces form a vector bundle
$R^Np_{N*}\CL$ over $X_n$ equipped with a flat Gauss-Manin connection.
In \cite{sv} certain maps compatible with the connections
\begin{equation}
\label{map-omega}
\omega: \CB(\Lambda_1,\ldots,\Lambda_{n+1})\lra
R^Np_{N*}\CL(\Lambda_1,\ldots,\Lambda_{n+1})^{\Sigma,-}
\end{equation}
where constructed (here the superscript $"\Sigma,-"$ denotes the subbundle
of skew invariants). The second main result of the present paper is
(see ~\ref{main}):

{\bf Theorem.} {\em The map $\omega$ passes through the projection
$\CB(\Lambda_1,\ldots,\Lambda_{n+1})\lra\CC(\Lambda_1,\ldots,\Lambda_{n+1})$
and thus induces the map
$$
\bomega: \CC(\Lambda_1,\ldots,\Lambda_{n+1})\lra
R^Np_{N*}\CL(\Lambda_1,\ldots,\Lambda_{n+1})^{\Sigma,-}
$$}

There are reasons to expect that the map $\bomega$ is injective.
It would be very interesting to define its image in topological terms;
if the above expectation is true, we would have a topological description
of the bundle of conformal blocks.

\subsection{} The paper goes as follows.
Section 1 is devoted to the proof of Theorem ~\ref{explicit}.

The main aim of Sections 2 and 3 is to define the map ~(\ref{map-omega}).
This map actually was introduced in ~\cite{sv2}, \cite{sv}. However,
{}~\cite{sv2} did not contain proofs, and the result ~\cite{sv} was formulated
in a greater generality, and we need here some important details
not formulated explicitely in {\em loc.cit.}
We include these details in Section 2. In this Section we discuss the
beautifull interrelation between certain spaces of rational functions
on affine spaces, graphs, and free Lie algebras.
We believe that the contents of this Section might be of independent interest.

At the end of Section 3 we formulate Theorem ~\ref{main}.

Sections 4 and 5 are devoted to the proof of Theorem ~\ref{main}.
This result is equivalent to the claim that certain differential
forms are exact. In Theorem ~\ref{main-expl} we write down certain
identity between differential forms, which is more general and precise than
the above claim. We call it {\bf Resonance identity}. In Section 5 we
prove it.

\subsection{} The results of this paper have been announced in ~\cite{fsva}.
The proof for the case $\fg=sl(2)$ is given in ~\cite{fsv1}.

Although the present paper heavily depends on the main construction
of ~\cite{sv}, we regard it as practically self-contained. In fact,
we tried to include into Sections 2 and 3 all the results
from {\em loc.cit.} which we need; the proofs are either given or
straightforward.
We hope that this alternative exposition is usefull also for a
better understanding of a more general framework of {\em loc.cit.}

We are greately indebted to Michael Finkelberg for his permission
to include his proof of the key point of Theorem ~\ref{explicit}.
Our initial proof was more complicated.


\section{Spaces of conformal blocks}
\label{conf}

\subsection{}
\label{notations} Throughout the paper we fix a complex finite dimensional
simple Lie algebra $\fg$ with a chosen system of Chevalley generators
$f_i,e_i,h_i,\ i=1,\ldots, r$. Let $\fg=\fn_-\oplus\fh\oplus\fn_+$
be the corresponding Cartan decomposition; $\alpha_1,\ldots,\alpha_r\in \fh^*$
the simple roots. (For a vector space $V$, $V^*$ will always denote
the dual vector space.)

Let $\omega:\fg\iso \fg$ denote {\em Chevalley involution} of $\fg$ -
the Lie algebra isomorphism such that $\omega(f_i)=-e_i,\
\omega(e_i)=-f_i,\ \omega(h_i)=-h_i$.

\begin{equation}
\label{hi-root}
\theta=\sum_{i=1}^ra_i\alpha_i
\end{equation}
will denote the highest root;
\begin{equation}
\label{du-hi-root}
\theta^{\vee}=\sum_{i=1}^ra^{\vee}_ih_i
\end{equation}
the highest root of the dual root system. Here all $a_i,a^{\vee}_i$ are
positive
integers. We set
$$
g=\sum_{i=1}^ra^{\vee}_i+1
$$
- it is the dual Coxeter number of $\fg$, ~\cite{k},\ 6.0.

Let
\begin{equation}
\label{nu-isom}
\nu:\fh\iso\fh^*
\end{equation}
denote the isomorphism such that $\nu(h_i)=a_i^{\vee-1}a_i\alpha_i$
(cf. ~\cite{k},\ 6.2.2).Let
$$
(,):\fh\times\fh\lra\Bbb C
$$
denote the corresponding bilinear form; it is non-degenerate and symmetric.
We denote by the same symbol the bilinear form on $\fh^*$ induced
by means of $\nu$. We have $(\theta,\theta)=2$. Evidently,
$\nu(\theta^{\vee})=\theta$.

We extend $(,)$ to the symmetric non-degenerate invariant bilinear form on
$\fg$
as in ~\cite{k}, ch. 2.
\begin{equation}
\label{omega-tens}
\Omega\in\fg\otimes\fg
\end{equation}
will denote the corresponding invariant symmetric tensor.

\subsection{} If $M$ is a representation of $\fh$, $\lambda\in\fh^*$, we set
$M_{\lambda}=\{ x\in M| hx=<\lambda,h>x\mbox{\ for all\ }h\in\fh\}$.
We will consider only {\em $\fh$-diagonalizable} representations of $\fg$, i.e.
such that $M=\oplus_{\lambda}M_{\lambda}$.

Set $M^0=\oplus_{\lambda}M^*_{\lambda}$; introduce an action of $\fg$ on $M^0$
by the formula $<cx^*,x>=<x^*,-\omega(c)x>$ for $x\in M,\ x^*\in M^0,\
c\in \fg$. This $\fg$-module is called {\em the contragradient} to $M$.

Given $\Lambda\in \fh^*$, $M(\Lambda)$ will denote the Verma module
over $\fg$, generated by the vacuum vector $v_{\Lambda}$ subject to
defining relations $\fn_+v_{\Lambda}=0,\ hv_{\Lambda}=<\Lambda,h>v_{\Lambda}$.
$L(\Lambda)$ will denote the unique maximal irreducible quotient of
$M(\Lambda)$; by abuse of the notations, we will denote by $v_{\Lambda}$
the image of $v_{\Lambda}$ in $L(\Lambda)$ too.

There is a unique $\fg$-module morphism
\begin{equation}
\label{shapo}
S: M(\Lambda)\lra M(\Lambda)^0
\end{equation}
such that $<S(v_{\Lambda}),v_{\Lambda}>=1$. We have
$L(\Lambda)=M(\Lambda)/\ker(S)$.

We can also consider $S$ as a bilinear form
$M(\Lambda)\times M(\Lambda)\lra \Bbb C$; it is called the
{\em Shapovalov form}.

The weight $\Lambda$ is called {\em dominant integral} if all $<\Lambda,h_i>$
are non-negative integers. This is equivalent to finite dimensionality
of $L(\Lambda)$.

Let $W$ denote the Weyl group of $\fg$, $w_0\in W$ the longest element.
For a dominant integral $\Lambda$, $w_0(\Lambda)$ is the lowest
weight of $L(\Lambda)$ (~\cite{b}, ch. VIII, \S 7, no. 2, Remark 2).
We shall denote
$$
\barLam=-w_0(\Lambda)
$$
It is again a dominant integral weight, and we have
$$
L(\barLam)=L(\Lambda)^*
$$

\subsection{} Let $T$ be an independent variable, $\Bbb C[[T]]$ the ring
of formal power series, $\Bbb C((T))$ the field of Laurent power series.
For $f(T),\ g(T)\in \Bbb C((T))$, introduce the notation
$$
\res_0(f(T)dg(T))=\mbox{\ coefficient at\ $T^{-1}$\ of\ $f(T)g'(T)$}.
$$

Set $\fg[[T]]=\fg\otimes_{\Bbb C}\Bbb C[[T]]\subset
\fg((T))=\fg\otimes_{\Bbb C}\Bbb C((T))$. These are Lie algebras with
the bracket $[c\otimes f(T),c'\otimes g(T)]=[c,c']\otimes f(T)g(T),\
c,c'\in\fg$.  Define $\hfg$ as a central extension of $\fg((T))$
$$
\hfg=\fg((T))\oplus\Bbb C\cdot\1
$$
where $\1$ lies in the centrum of $\hfg$, and
$$
[c\otimes f(T),c'\otimes g(T)]=[c,c']\otimes
f(T)g(T)+(c,c')\res_0(f(T)dg(T))\cdot\1
$$

Set
$$
\tfg=\fg[T,T^{-1}]\oplus\Bbb C\cdot\1
$$
It is a Lie subalgebra of $\hfg$.

We have natural embeddings $\fg\subset \tfg\subset\hfg$; we will identify
$\fg$ with its image in these algebras. We denote by $\tfg^+$
(resp., $\hfg^+$) the Lie subalgebra $\fg[T]\oplus \Bbb C\cdot\1\subset \tfg$
(resp., $\fg[[T]]\oplus \Bbb C\cdot\1\subset \hfg$).

Let us choose an element $e_{\theta}$ in the root subspace $\fg_{\theta}$
such that $(e_{\theta},-\omega(e_{\theta}))=1$; set $f_{\theta}=
\omega(e_{\theta})$. We have $[e_{\theta},f_{\theta}]=\theta^{\vee}$.

Set $e_0=f_{\theta}T,\ f_0=e_{\theta}T^{-1}$. The elements $e_0,\ldots,e_r,\
f_0,\ldots f_r$ form a system of generators of $\tfg$, ~\cite{k}, ch. 7.
We have
\begin{equation}
\label{e0f0}
[e_0,f_0]=\1-\theta^{\vee}
\end{equation}

\subsection{}
\label{reps} All representations $V$ of $\tfg$ we will consider will
have the following finiteness property: for every $x\in V$ there exists
$n\in \Bbb Z$  such that $cT^{n'}x=0$ for all $n'\geq n,\ c\in \fg$. For such
representations, the action of $\tfg$ may be extended uniquely to the
action of $\hfg$, cf. ~\cite{kl}, no. 1.

We fix a positive integer $k$; unless specified otherwise, $\1$ will
act as the multiplication by $k$ on all our representations.
We set $\kappa=k+g$.

Let $M$ be a representation of $\fg$. Consider $M$ as a $\tfg^+$-module,
by setting $T\fg[T]$ to act as zero, and $\1$ as $k$. Set
$$
\tM=U(\tfg)\otimes_{U(\tfg^+)}M
$$
This $\tfg$-module is called {\em the generalized Weyl module}
associated to $M$. We have a natural embedding $M\subset \tM$.

If $M=L(\Lambda)$, we denote $\tM$ by $\tL(\Lambda)$. This module
has a unique irreducible quotient which will be denoted by
$\bL(\Lambda)$. We have an embedding $L(\Lambda)\subset\bL(\Lambda)$.

A weight $\Lambda$ is called {\em small} if it is dominant integral
and $(\Lambda,\theta)\leq k$. The set of all small weights will
be denoted by $C\subset\fh^*$. Define $s_0:\fh^*\lra \fh^*$ as
$$
s_0(\lambda)=\lambda+(k-(\lambda,\theta)+1)\theta
$$

For $\Lambda\in C$, the irreducible module $\bL(\Lambda)$ is the quotient
of the Weyl module $\tL(\Lambda)$ by the $\tfg$-submodule $L'$
generated by the singular vector $f_0^{k-(\Lambda,\theta)+1}v_{\Lambda}$.
These irreducible modules will be most important in the sequel. We have
an exact sequence
\begin{equation}
\label{singul}
\tL(s_0(\Lambda))\lra \tL(\Lambda)\lra \bL(\Lambda)\lra 0
\end{equation}
where the first map sends $v_{s_0(\Lambda)}$ to
$f_0^{k-(\Lambda,\theta)+1}v_{\Lambda}$.

\subsection{Spaces of coinvariants} For a positive integer $n$ denote by
$\hfg^n$ the central extension of the $n$-th cartesian power $\fg[[T]]^n$
$$
\fg[[T]]^n\oplus\Bbb C\cdot \1
$$
with the bracket
$$
[(c_i\otimes f_i(T))_{1\leq i\leq n},(c'_i\otimes g_i(T))_{1\leq i\leq n}]=
([c_i,c'_i]\otimes f_i(T)g_i(T))_{1\leq i\leq n}\oplus
\sum_{i=1}^n(c_i,c'_i)\res_0(f(T)dg(T))\cdot\1
$$
(it is {\em not} the $n$-th cartesian power of $\hfg^n$). The Lie subalgebra
$\tfg^n\subset\hfg^n$ is defined analogously.

Consider the Riemann sphere $\Bbb P^1=\Bbb P^1(\Bbb C)$, with a fixed
coordinate $z$. Let us pick $n+1$ distinct points $z=z_i,\ i=1,\ldots, n+1$.
We suppose that $z_i\in \Bbb C$ for $1\leq i\leq n$, and $z_{i+1}=\infty$.

Let $\fg(z_1,\ldots,z_{n+1})$ denote by the Lie algebra of rational functions
on $\Bbb P^1$ with values in $\fg$, regular outside $z_1,\ldots, z_{n+1}$.
We have local coordinates at our punctures: $z-z_i$ for $1\leq i\leq n$,
and $1/z$ at $\infty$. The Laurent expansions at our points define an embedding
\begin{equation}
\label{global}
\fg(z_1,\ldots,z_{n+1})\hra \hfg^{n+1}
\end{equation}
(see for example ~\cite{kl}II, 9.9 or ~\cite{fsv1}, 2.3).

For a Lie algebra $\fa$ and an $\fa$-module $M$, we will denote by
$M_{\fa}$ the space of coinvariants $M/\fa M$.

Given $n+1$ $\hfg$-modules $M_1,\ldots, M_{n+1}$,
the algebra $\hfg^n$ acts naturally on the tensor product
$M_1\otimes\ldots\otimes M_{n+1}$ (recall that $\1$ acts as $k$ on each $M_i$
by our assumtion). Using ~(\ref{global}), we regard
this tensor product as a $\fg(z_1,\ldots,z_{n+1})$-module, and
can consider the space of coinvariants
$$
(M_1\otimes\ldots\otimes M_{n+1})_{\fg(z_1,\ldots,z_{n+1})}
$$

\subsection{}
\label{shapiro}
\begin{lem}{} Let $M_1,\ldots,M_{n+1}$ be $\fg$-modules.
The embeddings $M_i\hra \tM_i$ induce an isomorphism of coinvariants
$$
(M_1\otimes\ldots\otimes M_{n+1})_{\fg}\cong
(\tM_1\otimes\ldots\otimes \tM_{n+1})_{\fg(z_1,\ldots,z_{n+1})}
$$
\end{lem}

{\bf Proof.} See \cite{fsv1}, 2.3.1 or \cite{kl}II, 9.15. $\Box$

\subsection{}
Let $L(\Lambda)$ be a finite dimensional irreducible representation
of $\fg$. Pick a lowest vector $v^0\in L(\Lambda)$(i.e. such that $\fn_-v_0=0$;
it is unique up to a constant). We have $v^0\in L(\Lambda)_{w_0(\Lambda)}$.

Let $X$ be a finite dimensional representation of $\fg$.  Let us consider
the maps
$$
X_{\barLam}\hra X\otimes L(\Lambda)\lra (X\otimes L(\Lambda))_{\fg}
$$
where the first one sends $x$ to $x\otimes v^0$, and the second one is the
projection. They induce map
\begin{equation}
\label{lowest}
(X_{\fn_-})_{\barLam}\lra (X\otimes L(\Lambda))_{\fg}
\end{equation}
\subsubsection{}
\label{lowest-lem}
\begin{lem}{} The map ~(\ref{lowest}) is an isomorphism.
\end{lem}

{\bf Proof.} Easy. (Cf. ~\cite{fsv1}, 2.3.3). $\Box$

\subsection{}
\label{w} Let $\Lambda_1,\ldots,\Lambda_{n+1}$ be dominant integral
weights.

Introduce a notation
$$
W=(L(\Lambda_1)\otimes\ldots\otimes L(\Lambda_n))_{\fn_-}.
$$
As usually, $W_{\lambda}$ will denote a weight component.

\subsubsection{}
\begin{cor}{} The map
\begin{equation}
\label{weyl}
W_{\barLam_{n+1}}\lra (\tL(\Lambda_1)\otimes\ldots\otimes \tL(\Lambda_{n+1}))
_{\fg(z_1,\ldots,z_{n+1})}
\end{equation}
sending $x_1\otimes\ldots\otimes x_n$ to
$x_1\otimes\ldots\otimes x_n\otimes v^0_{n+1}$, where
$v^0_{n+1}\in L(\Lambda_{n+1})$ is a lowest vector,
is an isomorphism.
\end{cor}

{\bf Proof.} Follows from ~\ref{shapiro} and ~\ref{lowest-lem} applied
to $X=L(\Lambda_1)\otimes\ldots\otimes L(\Lambda_n)$. $\Box$

\subsection{}
\label{blocks} Suppose now that all $\Lambda_i\in C$. Set for brevity
$L=L(\Lambda_1)\otimes\ldots\otimes L(\Lambda_n)$. Consider the operator
$$
\bz\cdot f_{\theta}=\sum_{i=1}^nz_if^{(i)}_{\theta}:L\lra L,
$$
where $f^{(i)}_{\theta}$ acts as $f_{\theta}$ on the factor $L(\Lambda_i)$
and as the identity on other factors. It gives rise to  the operator
$$
(\bz\cdot f_{\theta})^{k-(\Lambda_{n+1},\theta)+1}:L_{s_0(\barLam_{n+1})}\lra
L_{\barLam_{n+1}}
$$
which induces the operator
$$
W_{s_0(\barLam_{n+1})}\lra W_{\barLam_{n+1}}
$$
Let us denote this operator by $\bT(\bz)$. We set
$$
W(z_1,\ldots,z_n)=W_{\barLam_{n+1}}/\Ima(\bT(\bz))
$$

\subsection{}
\label{explicit}
\begin{thm}{} Suppose that $\Lambda_1,\ldots,\Lambda_{n+1}\in C$. The
isomorphism ~(\ref{weyl}) induces an isomorphism
$$
W(z_1,\ldots,z_n)\cong (\bL(\Lambda_1)\otimes\ldots\otimes \bL(\Lambda_{n+1}))
_{\fg(z_1,\ldots,z_{n+1})}
$$
\end{thm}

The proof will follow some lemmas.

\subsection{}

Suppose we have $\Lambda\in C$. One sees easily that the weight
$s_0(\Lambda)$ is dominant integral. Let us choose lowest vectors
$v_{\Lambda}^0\in L(\Lambda)$ and $v_{s_0(\Lambda)}^0\in L(s_0(\Lambda))$.

\subsubsection{}
\begin{lem}{} One has an exact sequence
\begin{equation}
\label{dsingul}
\tL(s_0(\Lambda))\lra \tL(\Lambda)\lra \bL(\Lambda)\lra 0
\end{equation}
where the first map sends $v^0_{s_0(\Lambda)}$ to
$f_{\theta}^{k-(\Lambda,\theta)+1}v^0_{\Lambda}$.
\end{lem}

{\bf Proof.} There exists a unique involution
\begin{equation}
\label{tomega}
\tomega:\tfg\lra\tfg
\end{equation}
such that $\tomega(c)=\omega(c)$ for $c\in\fg$, and
$\tomega(f_0)=f_{\theta}T^{-1},\ \tomega(e_0)=e_{\theta}T$.

For a $\tfg$-module $M$, denote by $^{\tomega}M$ the $\tfg$-module
obtained from $M$ by the restriction of scalars using ~(\ref{tomega}).
We have an isomorphism
$$
^{\tomega}\tL(\Lambda)\iso\tL(\barLam)
$$
sending $v^0_{\Lambda}$ to $v_{\barLam}$.
(Note that $^{\omega}L(\Lambda)\cong L(\barLam)$).

Since $\theta$ is the highest weight of the
adjoint representation, $\bar{\theta}=\theta$. It follows from the
$W$-invariance of $(\ ,\ )$ that $\ol{s_0(\Lambda)}=s_0(\barLam)$.

Now, if we apply $^{\tomega}$ to ~(\ref{singul}), we get ~(\ref{dsingul}).
$\Box$

\subsection{}
\label{first}
\begin{cor}{} The isomorphism ~(\ref{weyl}) induces an isomorphism
$$
W(z_1,\ldots,z_n)\cong (\tL(\Lambda_1)\otimes\ldots\otimes\tL(\Lambda_n)
\otimes\bL(\Lambda_{n+1}))_{\fg(z_1,\ldots,z_{n+1})}
$$
\end{cor}

{\bf Proof.} Apply the functor $(L\otimes\cdot)_{\fg(z_1,\ldots,z_{n+1})}$ to
{}~(\ref{dsingul}). $\Box$

The rest of the argument is due to M.Finkelberg.

\subsection{}
\label{surj}
\begin{lem}{} Let $\Lambda\in C$, let $Y$ be the kernel of the projection
$\tL(\Lambda)\lra\bL(\Lambda)$. The operator $e_0=f_{\theta}T$ is surjective
on $Y$.
\end{lem}

{\bf Proof.} Let us denote by $\CO$ the category of $\tfg$-modules $M$ which
are:

(a) $\fh$-diagonalizable, and such that for all $\lambda\in \fh^*$,
$\dim(M_{\lambda})<\infty$. $\1$ acts as a multiplication by $k$.
(b) The subalgebra $\tfn^+=\fn^+\oplus T\fg[T]$ acts locally nilpotently
on $M$, i.e.
for every $x\in M,\ e\in\tfn^+$, $e^Nx=0$ for $N>>0$.

Denote by $\omega':\tfg\iso\tfg$ the Lie algebra involution
that sends $e_i$ to $-f_i$, and $f_i$ to $-e_i$ for $i=0,\ldots,r$.
Let us define the duality functor
$$
D:\CO\lra \CO
$$
as follows. For $M\in\CO$, consider the space
$M'=\oplus_{\lambda\in\fh}M^*_{\lambda}$ with the $\tfg$-action
$<xm^*,m>=<m^*,-\omega'(x)m>$. By definition, $D(M)\subset M'$ is the maximal
submodule on which $\tfn^+$ acts locally nilpotently.

$D$ is an exact contravariant functor, and $DD\cong\Id$.

Let us return to the lemma. Suppose that $e_0$ is not surjective
on $Y$. Then $f_0$ is not injective on $D(Y)$. Let $y\in \Ker(f_0)$.
Let $Z\subset D(Y)$ be the $\tfg$-submodule generated by $y$.
All operators $f_i,\ i=1,\ldots,r$, and $e_i,\ i=0,\ldots,r$, act
locally nilpotently on $D(Y)$, hence $Z$ is an integrable $\tfg$-module
(which means by definition that all $e_i,f_i$ are locally nilpotent
on it). Hence, $D(Z)$ is a non-zero integrable quotient
of $Y$. This contradicts to the fact that $\bL(\Lambda)$ is the maximal
integrable quotient of $\tL(\Lambda)$, (cf. \cite{k}, ch. 10). $\Box$

\subsection{} In the setup of our theorem, consider the tensor
product
$$
X=\tL(\Lambda_1)\otimes X_2\otimes\ldots\otimes X_n\otimes\bL(\Lambda_{n+1})
$$
where $X_i=\tL(\Lambda_i)$ or $\bL(\Lambda_i)$. As  usually, we consider $X$
as a $\fg(z_1,\ldots,z_{n+1})$-module. We have
$$
f_{\theta}Ty\otimes x\equiv -y\otimes Ax\mod \fg(z_1,\ldots,z_{n+1})X
$$
for $y\in \tL(\Lambda_1),\ x\in X':= X_2\otimes\ldots\otimes
X_n\otimes\bL(\Lambda_{n+1})$, where $A$ is a linear operator on $X'$ which
acts as
$$
A(x_2\otimes\ldots\otimes x_{n+1})=
u_1f_{\theta}x_2\otimes\ldots\otimes x_{n+1}+\ldots +
x_2\otimes\ldots\otimes u_nf_{\theta}x_n\otimes x_{n+1} +
x_2\otimes\ldots\otimes x_n\otimes f_{\theta}T^{-1}x_{n+1}
$$
for some $u_i\in\Bbb C$.

\subsubsection{}
\label{nilp}
\begin{lem}{} $A$ is locally nilpotent on $X'$.
\end{lem}

In fact, it is easy to see that $f_{\theta}$ is locally
nilpotent on all $X_i$, and $f_{\theta}T^{-1}$ is locally
nilpotent on $\bL(\Lambda_{n+1})$ since this module is $\tfg$-integrable
(cf. \cite{tk}, 1.4.6). Hence, $A$ is locally nilpotent on $X'$. $\Box$

Let $Y_1=\ker(\tL(\Lambda_1)\lra \bL(\Lambda_1))$. We have an exact
sequence
$$
(Y_1\otimes X')_{\fg(z_1,\ldots,z_{n+1})}\lra
(\tL(\Lambda_1)\otimes X')_{\fg(z_1,\ldots,z_{n+1})}\overset{\phi}{\lra}
(\bL(\Lambda_1)\otimes X')_{\fg(z_1,\ldots,z_{n+1})}\lra 0.
$$
Hence, any element in $\ker(\phi)$ is of the form $\sum_jy_j\otimes x_j$
with $y_j\in Y_1,\ x_j\in X'$. It follows from ~\ref{surj} and ~\ref{nilp}
that such an element must be zero.

It follows that
$$
\phi:
(\tL(\Lambda_1)\otimes X_2\otimes\ldots\otimes X_n\otimes \bL(\Lambda_{n+1}))
_{\fg(z_1,\ldots,z_{n+1})} \lra
(\bL(\Lambda_1)\otimes X_2\otimes\ldots\otimes X_n\otimes \bL(\Lambda_{n+1}))
_{\fg(z_1,\ldots,z_{n+1})}
$$
is an isomorphism. Applying the same argument to other factors $X_i$
instead of $X_1$, we get the statement of Theorem~\ref{explicit} from
\ref{first}. $\Box$

\section{Trees, rational functions and Lie algebras}
\label{trees}

\subsection{} Let us fix a finite set $\CI$ and a set of distinct
complex numbers $\bz=\{ z_1,\ldots, z_n\}$, $n\geq 1$.

In the sequel, given a positive integer $m$, we will use the notation
$[m]:=\{ 1,\ldots, m\}$.

For every subset $I\subset \CI$ denote $\tI:= I\cup [n]$.

Let us consider finite oriented graphs $\Gamma$ whose set of vertices
$\Ver (\Gamma)$ is identified with a subset of $\tCI$.
We will denote by $\Ar(\Gamma)$ the set of arrows of $\Gamma$, and
$$
b,e:\Ar(\Gamma)\overset{\lra}{\lra}\Ver(\Gamma)
$$
the source and target of arrows respectively. We will call
{\em a support of $\Gamma$} the subset
$$
\Supp(\Gamma)=b(\Ar(\Gamma))\cup e(\Ar(\Gamma))\subset\Ver(\Gamma)\subset \tCI.
$$
The vertex $v\in\Ver(\Gamma)$ lying outside $\Supp(\Gamma)$ is called
{\em isolated}.

We will suppose that:

(a) every pair of vertices is joined by not more than one arrow;

(b) $\Gamma$ contains no loops.

Thus, $\Gamma$ is a disjoint union of trees. We will draw $\Gamma$ as
a graph with vertices labeled by elements $i\in \CI$ or $s\in [n]$,
and we will not picture isolated vertices.

We will denote the set of all such graphs by $Gr$, and call its elements
simply "graphs".

For a subset $I\subset \CI,\ s\in [n]$, $Gr_I$ (resp., $Gr_{I;s}$) will
denote the subset of all {\em connected} graphs with support equal to $I$
(resp., $I\cup\{ s\}$).

\subsection{} Let us define a $\Bbb C$-vector space $\fG$ with generators
$[\Gamma]$, $\Gamma\in Gr$, subject to the following relations.

\subsubsection{} If $\Gamma'$ is obtained from $\Gamma$ by removing
an isolated vertex then $[\Gamma']=[\Gamma]$.

\subsubsection{} If $\Gamma'$ is obtained from $\Gamma$ by
reversing one arrow, then $[\Gamma']=-[\Gamma]$.

\subsubsection{Triangle relation} Let $\{ i,j,k\}\subset \CI$ be any
three-element subset.
Consider three graphs:


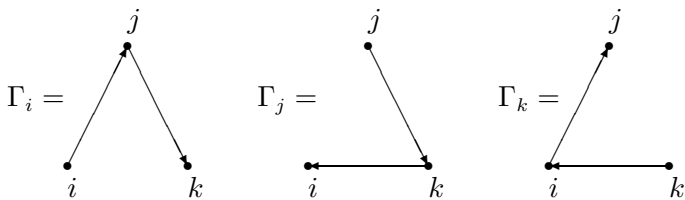
\begin{picture}(15,3)

\put(0,1.5){$\Gamma_i=$}

\put(1,0.5){\circle*{0.15}}
\put(1,0){$i$}
\put(2,2.5){\circle*{0.15}}
\put(2,2.8){$j$}
\put(3,0.5){\circle*{0.15}}
\put(3,0){$k$}

\put(1,0.5){\vector(1,2){1}}
\put(2,2.5){\vector(1,-2){1}}

\put(4,1.5){$\ \Gamma_j=$}

\put(5,0.5){\circle*{0.15}}
\put(5,0){$i$}
\put(6,2.5){\circle*{0.15}}
\put(6,2.8){$j$}
\put(7,0.5){\circle*{0.15}}
\put(7,0){$k$}

\put(6,2.5){\vector(1,-2){1}}
\put(7,0.5){\vector(-1,0){2}}

\put(8,1.5){\ $\Gamma_k=$}

\put(9,0.5){\circle*{0.15}}
\put(9,0){$i$}
\put(10,2.5){\circle*{0.15}}
\put(10,2.8){$j$}
\put(11,0.5){\circle*{0.15}}
\put(11,0){$k$}

\put(9,0.5){\vector(1,2){1}}
\put(11,0.5){\vector(-1,0){2}}


\end{picture}

Let $\Gamma$ be a graph such that all three graphs $\Gamma_i\cup\Gamma$ belong
to $Gr$. Then
$$
[\Gamma_i\cup\Gamma]+[\Gamma_j\cup\Gamma]+[\Gamma_k\cup\Gamma]=0
$$

\subsubsection{} If $\Gamma'$ is obtained from $\Gamma$ by removing
an arrow which begins at $s$ and ends at $s'$ where $s,s'\in [n]$, then
$$
[\Gamma']=(z_s-z_{s'})[\Gamma]
$$
\\

\subsubsection{Compositions} For $I\subset\CI$, we will denote by
$\fG_I,\ \fG_{I;s}$ the subspace of
$\fG$ generated by all $[\Gamma]$ where $\Gamma\in Gr_I$,
$\Gamma\in Gr_{I;s}$ respectively.

Let us denote by $\fG_{\tI}$ the space generated by all $[\Gamma]$ where
$\Supp(\Gamma)\subset\tI$, and each connected component of $\Supp(\Gamma)$
contains an element $s\in [n]$.

We have operations of partial multiplications
$$
\cdot: \fG_I\otimes\fG_J\lra \fG,
$$
$$
\cdot: \fG_{I;s}\otimes\fG_{J;s'}\lra \fG,
$$
$[\Gamma]\otimes [\Gamma']\mapsto [\Gamma]\cdot
[\Gamma']:=[\Gamma\cup\Gamma']$, defined iff $\card(I\cap J)\leq 1$.
They are associative and commutative in the obvious sense.

If $\card(I\cap J)=1$ then $\fG_I\cdot\fG_J\subset\fG_{I\cup J}$.
If $I\cap J=\emptyset$ then $\fG_{I;s}\cdot\fG_{J;s}\subset \fG_{I\cup J; s}$.

For $I\subset \CI$ denote by $Seq(I)$ the set of all sequences
\begin{equation}
\label{seq}
\vI=(i_1,\ldots,i_N)
\end{equation}
such that $I=\{ i_1,\ldots, i_N\}$. In other words, $Seq(I)$ is the set
of all total orders on $I$. Evidently, $Seq(I)$ is an $\Aut(I)$-torsor.

Notational remark: if $\vI$ is a sequence as above, we denote by $I$ the
set of its elements.

We define the graphs:


\begin{picture}(15,3)

\put(0,1.5){$\Gamma_{\vI}=$}

\put(2,1.5){\circle*{0.15}}
\put(2,1){$i_N$}
\put(2,1.5){\vector(1,0){2}}
\put(4,1.5){\circle*{0.15}}
\put(4,1){$i_{N-1}$}
\put(4,1.5){\vector(1,0){2}}

\put(7,1.5){$\ldots$}

\put(8,1.5){\vector(1,0){2}}
\put(10,1.5){\circle*{0.15}}
\put(10,1){$i_2$}
\put(10,1.5){\vector(1,0){2}}
\put(12,1.5){\circle*{0.15}}
\put(12,1){$i_1$}

\end{picture}

and


\begin{picture}(15,3)

\put(0,1.5){$\Gamma_{\vI; s}=$}

\put(2,1.5){\circle*{0.15}}
\put(2,1){$i_N$}
\put(2,1.5){\vector(1,0){2}}
\put(4,1.5){\circle*{0.15}}
\put(4,1){$i_{N-1}$}
\put(4,1.5){\vector(1,0){2}}

\put(7,1.5){$\ldots$}

\put(8,1.5){\vector(1,0){2}}
\put(10,1.5){\circle*{0.15}}
\put(10,1){$i_2$}
\put(10,1.5){\vector(1,0){2}}
\put(10,1.5){\circle*{0.15}}
\put(12,1){$i_1$}
\put(12,1.5){\vector(1,0){2}}
\put(14,1.5){\circle*{0.15}}
\put(14,1){$s$}

\end{picture}

for $s\in [n]$.

The following two lemmas are checked directly.

\subsubsection{}
\begin{lem}{} For a given $I\subset\CI,\ s\in [n]$, elements
$[\Gamma_{\vI; s}]$, $\vI\in Seq(I)$, form a basis of $\fG_{I;s}$. $\Box$
\end{lem}

\subsubsection{}
\begin{lem}{} Let us pick an element $i_0\in I$. All elements
$[\Gamma_{\vI}]$ where $\vI\in Seq (I)$ is such that its first element
is $i_0$, form a basis of $\fG_I$. $\Box$
\end{lem}

\subsection{} Let us consider a field $\CF=\Bbb C(\bt)$ of rational
functions of the set $\bt=\{ t_i\}_{i\in \CI}$ of commuting
independent variables indexed by $\CI$. For a subset $I\subset \CI$ denote
$\bt_I:=\{ t_i\}_{i\in I}\subset \bt$.

Let us assign to each $\Gamma\in Gr$ a rational function
$$
E(\Gamma)=\prod_{a\in\Ar(\Gamma)}\frac{1}{t_{b(a)}-t_{e(a)}}
$$
where we agree that $t_s:=z_s$ for $s\in [n]$.
One sees easily that one gets a well defined map
$$
E:\fG\lra \CF
$$
Moreover, $E$ is compatible with multiplication in the evident sense.

For $I\subset \CI$, let us define subspaces $R_I\subset R_I(\bz)\subset \CF$ as
$R_I=E(\fG_I),\ R_I(\bz)=E(\fG_{\tI})$.

\subsubsection{}
\begin{lem}{} The map $E$ induces isomorphisms
$$
E: \fG_I\iso R_I;\ E:\fG_{\tI}\iso R_I(\bz).\ \Box
$$
\end{lem}

For $\vI$ as in ~(\ref{seq}) will use the notations
$$
E_{\vI}(\bt_I)=E(\Gamma_{\vI})=\prod_{j=2}^N\frac{1}{t_{i_j}-t_{i_{j-1}}}
$$
$$
E_{\vI;s}(\bt_I;z_s)=E(\Gamma_{\vI;s})=\frac{1}{t_{i_1}-z_s}\cdot
\prod_{j=2}^N\frac{1}{t_{i_j}-t_{i_{j-1}}}
$$

\subsection{Shapovalov form} Suppose we are given a symmetric map
$$
a: \CI\times\CI\lra \Bbb C
$$
and numbers $a(i,s)\in \Bbb C$ for each $i\in\CI,\ s\in [n]$.

(i) Given two sequences $\vI =(i_0,\ldots,i_N),\ \vJ =(j_0,\ldots,j_N)\in
\CI^{N+1}\ (N\geq 0)$ such that $i_0=j_0$, define a complex number $S(\vI,\vJ)$
by induction on
$N$. Namely, set $S((i),(i))=1$, and
$$
S(\vI,\vJ)=\sum_{q\in [N]: i_q=j_N}(\sum_{p=0}^qa(i_p,j_N))\cdot
S((i_0,\ldots,\hat{i}_p,\ldots,i_N),(j_0,\ldots,j_{N-1}))
$$
for $N\geq 1$.

It is clear that $S(\vI,\vJ)\not=0$ only if $I=J$.
One easily proves that
$$
S(\vI,\vJ)=S(\vJ,\vI)
$$

(ii) Given arbitrary $\vI =(i_1,\ldots,i_N),\ \vJ =(j_1,\ldots,j_N)\in
\CI^{N}\ (N\geq 0)$ and $s\in [n]$, define a complex number $S(\vI,\vJ)^{(s)}$
by induction on $N$. Namely, set $S((),())^{(s)}=1$, and
$$
S(\vI,\vJ)^{(s)}=\sum_{q\in [N]: i_q=j_N}(a(s,j_N)+\sum_{p=1}^qa(i_p,j_N))\cdot
S((i_1,\ldots,\hat{i}_p,\ldots,i_N),(j_1,\ldots,j_{N-1}))^{(s)}
$$
for $N\geq 1$.

Again $S(\vI,\vJ)^{(s)}\not=0$ only if $I=J$, and
$$
S(\vI,\vJ)^{(s)}=S(\vJ,\vI)^{(s)}
$$

\subsection{}
Given $I\subset \CI$, $\vI=(i_1,\ldots,i_N)\in Seq(I)$ and $s\in [n]$,
define important rational functions
\begin{equation}
\label{a-fun}
A_{\vI}(\bt_I)=\prod_{p=2}^N(\sum_{q=1}^{p-1}\frac{a(p,q)}{t_{i_p}-t_{i_q}})
\end{equation}
and
\begin{equation}
\label{b-fun}
B_{\vI}(\bt_I;z_s)=\prod_{p=1}^N
(\frac{a(p,s)}{t_{i_p}-z_s}+\sum_{q=1}^{p-1}\frac{a(p,q)}{t_{i_p}-t_{i_q}})
\end{equation}

\subsubsection{}
\label{key}
{\bf Key Lemma.\ }{\em We have
$$
B_{\vI}(\bt;z_s)=\sum_{\vJ=(j_1,\ldots,j_N)\in Seq(I)}S(\vI,\vJ)^{(s)}
\cdot E_{\vI}(\bt_I;z_s)\
$$}

{\bf Proof.} This is a particular case of ~\cite{sv}, Thm. 6.6
(for $\cdot =0$ and $\lambda=(1,\ldots, 1)$). Or else, it may be checked
directly. $\Box$

\subsection{} Let $\fn=\Lie (f_i)_{i\in \CI}$ be a free Lie algebra
on generators $f_i$, $i\in \CI$ (over $\Bbb C$). Its enveloping algebra
$U(\fn)$ may be identified with a free associative $\Bbb C$-algebra with
these generators.

For a sequence
\begin{equation}
\label{seqrep}
\vI=(i_1,\ldots,i_N)\in\CI^N
\end{equation}
denote by $f_{\vI}\in U(\fn)$ the monomial
$$
f_{\vI}=f_{i_N}f_{i_{N-1}}\cdot\ldots\cdot f_{i_1},
$$
and by $[f_{\vI}]\in\fn$ the commutator
$$
[f_{\vI}]=\ad(f_{i_N})\circ\ad(f_{i_{N-1}})\circ\ldots\circ
\ad(f_{i_2})(f_{i_1}),
$$
where $\ad(x)(y):=[x,y]$. (Note the reverse order!)

Given $I\subset \CI$, let $U(\fn)_I\in U(\fn)$ (resp., $\fn_I\in\fn$) be
the $\Bbb C$-subspace generated by all monomials $f_{\vI}$ (resp., by
all commutators $[f_{\vI}]$), $\vI\in Seq(I)$. We set
$U(\fn)_{\emptyset}=\Bbb C\cdot 1;\ \fn_{\emptyset}= 0.$

Let us denote by $Seq^{(n)}(I)$
the set of all $n$-tuples of sequences $\vI_1,\ldots,\vI_n$ such that $I$
is a {\em disjoint} union $\cup_{s=1}^n I_s$.

We denote by $U(\fn)^{\otimes n}_I\subset U(\fn)^{\otimes n}$
the $\Bbb C$-subspace generated by all monomials of the form
$$
x=f_{\vI_1}\otimes\ldots\otimes f_{\vI_n}
$$
with $(\vI_1,\ldots,\vI_n)\in Seq^{(n)}(I)$.

\subsection{} Suppose we have  $(\vI_1,\ldots\vI_n)\in Seq^{(n)}(I)$
as above. We set
\begin{equation}
\label{BZ}
B_{\vI_1,\ldots,\vI_n}(\bt;\bz)=B_{\vI_1}(\bt_{I_1};z_1)
\cdot\ldots\cdot B_{\vI_n}({\bt_{I_n}; z_n})
\end{equation}

\subsection{}
\label{BA}
\begin{lem}{} Pick $I\subset \CI$.

(i) The assignment $f_{\vI}\mapsto B_{\vI}(\bt_I;z_s)$
defines the map
$$
B_I(z_s):\ U(\fn)_I\lra R_I(\bz).
$$

(ii) The assignment
$$
x=f_{\vI_1}\otimes\ldots\otimes f_{\vI_n}\mapsto
B_{(\vI_1,\ldots,\vI_n)}(\bt;\bz)
$$
defines the map
$$
B_I(\bz):U(\fn)^{\otimes n}_I\lra R_I(\bz).
$$
\end{lem}

{\bf Proof.} This is trivial since the above monomials form bases of the
left hand sides. $\Box$

\subsection{}
\label{A/B}
\begin{lem}{} We have
$$
B_I(z_s)([f_{\vI}])=A_{\vI}(\bt_I)\cdot
\sum_{i\in I}\frac{a(i,s)}{t_i-z_s}
$$
\end{lem}

{\bf Proof.} The lemma is proved by induction on $\card(I)$, simultaneously
with the following statement which is a particular case of ~\ref{A-B} below.

\begin{claim}{} Suppose that $j\not\in I$. Then
$$
B_{I\cup \{ j\}}(z_s)([f_{\vI}]f_j)=\frac{a(j,s)}{t_j-z_s}\cdot
A_{\vI}(\bt_I)\cdot \sum_{i\in I}
(\frac{a(i,j)}{t_i-t_j}+\frac{a(i,s)}{t_i-z_s}).
$$
\end{claim} $\Box$

\subsection{} Suppose we have $n+1$ disjoint subsets
$I_1,\ldots, I_n, J\subset \CI$, and sequences $\vI_j\in Seq(I_j),\ j\in [n];
\ \vJ\in Seq(J)$. Set $I=\cup_{j=1}^n I_j$.

For $x=x_1\otimes\ldots\otimes x_n\in U(\fn)^{\otimes n},\ s\in [n],\
y\in U(\fn)$, denote
\begin{equation}
\label{local}
y^{(s)}x=x_1\otimes\ldots\otimes yx_s\otimes\ldots\otimes x_n
\end{equation}

\subsubsection{}
\label{A-B}
\begin{lem}{} We have
$$
B_{I\cup J}(\bz)([f_{\vJ}]^{(s)}\cdot f_{\vI_1}\otimes\ldots\otimes f_{\vI_n})
= B_{\vI_1,\ldots,\vI_n}(\bt;\bz)\cdot A_{\vJ}(\bt_J)\cdot
(\sum_{i\in I_s;j\in J}\frac{a(j,i)}{t_j-t_i} +
 \sum_{j\in J}\frac{a(j,s)}{t_j-z_s})
$$
\end{lem}

{\bf Proof.} Similar to the proof of Lemma ~\ref{A/B} above, by induction on
$\card(I)$. $\Box$

\subsection{}
\label{A-map}
\begin{lem}{} The assignment $[f_{\vI}]\mapsto A_{\vI}(\bt_I)$ defines the map
$$
A_I:\ \fn_I\lra R_I.
$$
\end{lem}

{\bf Proof.} Follows from ~\ref{A/B}. $\Box$

\subsection{}
\label{A-A}
\begin{lem}{} Suppose we have two disjoint subsets $I,J\subset \CI$, and
$\vI\in Seq(I),\ \vJ\in Seq(J)$. Then
$$
A_{I\cup J}([[f_{\vI}],[f_{\vJ}]])=A_I([f_{\vI}])\cdot A_J([f_{\vJ}])
\cdot \sum_{i\in I; j\in J}\frac{a(j,i)}{t_j-t_j}
$$
\end{lem}

{\bf Proof.} Induction on $\card(J)$ using ~\ref{A-B} for $n=1$. $\Box$

\subsection{}
\label{unfold} We will need a generalization of the previous constructions
to the following situation. Let
$$
\pi: \CI\lra \CJ
$$
be a surjective map between two finite sets. We set
$$
k_j=\card (\pi^{-1}(j));\ \bk=(k_j)_{j\in \CJ}
$$
One can regard the pair $(\CJ,\bk)$ as a "weighted set", $k_j$ being
"the multiplicity" of $j$. We will say that the map $\pi$ is
{\em an unfolding} of $(\CJ,\bk)$.

We set $\Sigma_{\CI}:=\Aut (\CI)$ and
\begin{equation}
\label{sigma-pi}
\Sigma_{\pi}:=\prod_{j\in \CJ}\Aut (\pi^{-1}(j))\subset \Sigma_{\CI}
\end{equation}
the last group is non-canonically isomorphic to the product of symmetric groups
$\Sigma_{k_j}$.

We consider a Lie algebra $\fn=\Lie (f_j)_{j\in \CJ}$, and the
field $\CF =\Bbb C (t_i)_{i\in \CI}$, with subspaces $R_I,\ R_I(\bz),\
(I\subset \CI)$ as above.

The group $\Sigma_{\CI}$ acts on $\CF$ permuting generators $t_i$.
For every $J\subset \CJ$ we have an induced action of $\Sigma_{\pi}$ on
$R_{\pi^{-1}(J)},\ R_{\pi^{-1}(J)}(\bz)$. We will denote by $\Sym_{\pi}$ the
symmetrisation operator
$$
\Sym_{\pi}=\sum_{\sigma\in\Sigma_{\pi}}\sigma
$$
acting on these rings.

Pick $J\subset \CJ$; set $I=\pi^{-1}(J)$. Set $N:=\card (I)$; evidently
$N=\sum_{j\in J} k_j$.

Let us denote by $Seq(J;\bk)$ the set of all sequences
$$
\vJ=(j_1,\ldots,j_N),\ j_i\in J,
$$
such that for each $j\in J$ there are exactly $k_j$ entries equal to $j$
in $\vJ$. We will denote by $U(\fn)_{\CJ,\bk}$ (resp., $\fn_{\CJ,\bk}$) the
$\Bbb C$-space generated by all monomials $f_{\vJ}$ (resp.,
commutators $[f_{\vJ}]$) with $\vJ\in Seq(J;\bk)$.

For a sequence $\vI=\{ i_1,\ldots,i_N\}$ we write
$\pi(\vI)=(\pi(i_1),\ldots,\pi(i_N)\}$.

Given $\vJ\in Seq(J;\bk)$, let us call {\em an unfolding of $\vJ$
with respect to $\pi$} a sequence
$$
\vI=(i_1,\ldots,i_N)\in \CI^N,
$$
such that $\pi(\vI)=\vJ$ and all $i_p$ are distinct.
We denote the set of all unfoldings by $Unf(\vJ)$.It is naturally a torsor over
the group $\prod_{j\in J}\Aut (\pi^{-1}(j))$\footnote{Recall that a torsor
over a group $G$ is set $X$ equipped with a free and transitive action of $G$}.

Suppose we are given a symmetric map $\CI\times\CI\lra \Bbb C$.
Let us pick $\vI\in Unf(\vJ)$. Let us define rational functions
\begin{equation}
\label{AA}
A_{\vJ;\pi}(\bt_I)= Sym_{\pi}\{ A_{\vI}(\bt_I)\}\in R_I
\end{equation}
\begin{equation}
\label{BB}
B_{\vJ;\pi}(\bt_I;z_s)= Sym_{\pi}\{ B_{\vI}(\bt_I;z_s)\}\in R_I(\bz)
\end{equation}
where the functions in figure brackets are defined in ~(\ref{a-fun}),
{}~(\ref{b-fun}).

Analogously, for any positive integer $n$, denote by
$Seq^{(n)}(J;\bk)$ the set of all $n$-tuples of sequences
$\vJ_1,\ldots,\vJ_n$ such that their concatenation
$\vJ=\vJ_1|\ldots |\vJ_n$ belongs to $Seq(J;\bk)$. We will denote by
$U(\fn)^{\otimes n}_{J,\bk}\subset U(\fn)$ the subspace spanned by all
monomials
$$
x=f_{\vJ_1}\otimes\ldots\otimes f_{\vJ_n},\
(\vJ_1,\ldots,\vJ_n)\in Seq^{(n)}(J;\bk).
$$

Given such $(\vJ_1,\ldots,\vJ_n)$, we will call its {\em unfolding}
an $n$-tuple of sequences $(\vI_1,\ldots,\vI_n)$ such that for all
$s$ $\vI_s$ is an unfolding of $\vJ_s$, and all these sequences are
{\em disjoint}, i.e. the corresponding sets $I_1,\ldots,I_n$ do not
intersect. The set of unfoldings will be denoted
$Unf(\vJ_1,\ldots,\vJ_n)$.

Pick an unfolding $(\vI_1,\ldots,\vI_n)$. Define
\begin{equation}
\label{BBZ}
B_{\vJ_1,\ldots,\vJ_n;\pi}(\bt;\bz)=\Sym_{\pi}\{ B_{\vI_1}(\bt_{I_1};z_1)
\cdot\ldots\cdot B_{\vI_n}(\bt_{I_n};z_n)\}
\end{equation}

One can see that the above functions do not depend on a particular choice
of unfoldings, as the notation suggests.

\subsection{}
\begin{lem}{} (i) The assignment
$$
f_{\vJ}\mapsto B_{\vJ;\pi}(\bt_I;z_s)
$$
defines the map
$$
B_{J;\bk}(z_s): U(\fn)_{J;\bk}\lra R_I(\bz)^{\Sigma_{\pi}}.
$$

(ii) The assignment
$$
x=f_{\vJ_1}\otimes\ldots\otimes f_{\vJ_n}\mapsto
B_{\vJ_1,\ldots,\vJ_n;\pi}(\bt;\bz)
$$
defines the map
$$
B_{J;\bk}(\bz): U(\fn)^{\otimes n}_{J;\bk}\lra R_I(\bz)^{\Sigma_{\pi}}.
$$

(iii) The assignment
$$
[f_{\vJ}]\mapsto A_{\vJ;\pi}(\bt_I)
$$
defines the map
$$
A_{J;\bk}: \fn_{J;\bk}\lra R_I^{\Sigma_{\pi}}.
$$
\end{lem}

{\bf Proof} follows from the non-symmetrized case (Lemmas ~\ref{BA} and
{}~\ref{A-map} above). Cf. also ~\cite{sv}, 5.11. $\Box$


\section{Conformal blocks and de Rham cohomology}

\subsection{}

\subsubsection{} Let us introduce some notations.
For $\lambda\in\fh^*,\
\lambda=\sum_{i=1}^rq_i\alpha_i$, set $|\lambda|=\sum_{i=1}^rq_i$.

For $\lambda,\ \lambda'\in\fh^*$, we write $\lambda\leq\lambda'$ iff
$\lambda'-\lambda=\sum_{i=1}^rq_i\alpha_i$ where all $q_i$ are non-negative
integers.

\subsubsection{}
\label{number} Let us fix weights $\Lambda_1,\ldots,\Lambda_n\in\fh^*$;
set $\Lambda=\sum_{i=1}^n\Lambda_i$. Fix non-negative integers
$k_1,\ldots,k_r$, and set $\CJ=\{ i\in [r]|k_i>0\}$; $\bk=(k_j)_{j\in \CJ}$.
Set $N=\sum_{i=1}^rk_i,\ \alpha=\sum_{i=1}^rk_i\alpha_i,\
\Lambda'=\Lambda-\alpha$.

Let us pick an unfolding of $(\CJ,\bk)$:
\begin{equation}
\label{pi}
\pi: [N]\lra \CJ,
\end{equation}
where $\card (\pi^{-1}(j))=k_j$ for all $j\in\CJ$. We denote $\CI:=[N]$.

As in ~\ref{unfold}, one defines the symmetric group
$\Sigma:=\Sigma_{\pi}\subset \Sigma_{\CI}=\Sigma_N$.

Recall that we have fixed a positive integer $k$, and we set $\kappa=k+g$
(see ~\ref{reps}).

\subsubsection{} Let us consider the cartesian product of $N$ projective
lines, $X=(\Bbb P^1)^N$, with coordinates $(t_1,\ldots,t_N),\ t_i\in\Bbb C\cup
\{ \infty\}$. Fix $n$ distinct complex numbers $z_1,\ldots,z_n$,
and set $z_{n+1}=\infty$. Inside $X$, consider the following hyperplanes:
$$
H_{ij}:\ t_i=t_j,\ i,j=1,\ldots,N\ \mbox{(so, $H_{ij}=H_{ji}$)};\
H_{i;s}:\ t_i=z_s,\ i=1,\ldots,N;\ s=1,\ldots, n+1.
$$
We denote by $\bCC$ the set of all these hyperplanes. We set $\CC_{\infty}=
\{H_{i,n+1}\}_{i=1,\ldots,N}$, $\CC=\bCC -\CC_{\infty}$.

Let us define the map $a:\bCC\lra\Bbb C$ as follows. Set
$$
a(H_{ij})=(\alpha_{\pi(i)},\alpha_{\pi(j)})/\kappa;\
a(H_{i;s})=-(\Lambda_s,\alpha_{\pi(i)})/\kappa
$$
if $i<n+1$. Finally, set
$$
a(H_{i;n+1})=(\Lambda-\sum_{j\neq i}\alpha_{\pi(j)},\alpha_{\pi(i)})/\kappa
$$
We will also use the notations
$$
a(i,j)=a(H_{ij});\ a(i,s)=a(H_{i;s}).
$$
These numbers are determined by the condition that for every line
$L\cong\Bbb P^1\hra X$ defined by equations $t_j=z_{p_j},\ j\neq i,$
($i$ being fixed), the sum $\sum a(H)$ over all $H\in\bCC$ meeting
$L$ transversally, equals $0$.

Set $U=X-\cup_{H\in\bCC}H\subset X$. We will identify
$X-\cup_{H\in\CC_{\infty}}H$ with the affine space $\Bbb A^N$ with coordinates
$t_1,\ldots,t_N$. For each $H\in \CC$, we define the function $f_H$
on $\Bbb A^N$ as follows: $f_{H_{ij}}=t_i-t_j;\ f_{H_{i;s}}=t_i-z_s$.

Let us define the following complex
of vector spaces
\begin{equation}
\label{Omega}
\Omega^{\bullet}:0\lra\Omega^0\overset{d}{\lra}\ldots\overset{d}{\lra}\Omega^N\lra 0
\end{equation}
By definition, $\Omega^i$ is the space of holomorphic $i$-forms on $U$.
The differential $d$ is the sum
\begin{equation}
\label{nabla}
d=d_{DR}+\omega_a
\end{equation}
where $d_{DR}$ is the de Rham differential and $\omega_a$ denotes
the left multiplication by the closed $1$-form
$$
\omega_a=\sum_{H\in\CC}a(H)\dlog(f_H)
$$
where $\dlog(f_H)=d(f_H)/f_H$. We will write elements of $\Omega^i$
symbolically as
\begin{equation}
\label{form}
f(t_1,\ldots,t_N)\cdot l\cdot dt_{p_1}\wedge\ldots\wedge dt_{p_i}
\end{equation}
where $f$ is a holomorphic function, and
$$
l=l(t_1,\ldots,t_N)=\prod_{i,s}(t_i-z_s)^{a(i,s)}\prod_{i<j}(t_i-t_j)^{a(i,j)}
$$
- this expression should be considered as a formal symbol.
The formal differentiation of ~(\ref{form}) gives the differential
{}~(\ref{nabla}) since $\dlog (l)=\omega_a$.

The symmetric group $\Sigma$ acts on $\Omega^{\bullet}$ by the rule
\begin{equation}
\label{action}
\sigma(f(t_1,\ldots,t_N)\cdot l\cdot dt_{p_1}\wedge\ldots\wedge dt_{p_i})=
f(t_{\sigma(1)},\ldots,t_{\sigma(N)})\cdot l\cdot
dt_{\sigma(p_1)}\wedge\ldots\wedge dt_{\sigma(p_i)}
\end{equation}

The geometric meaning of $\Omega^{\bullet}$ is as follows. The form
$\omega_a$ defines an integrable connection $\nabla=d_{DR}+\omega_a$
on the sheaf $\CO_U$ of holomorphic functions on $U$.
$\Omega^{\bullet}$ is the complex of global sections of the holomorphic de Rham
complex associated with $\nabla$. It computes the cohomology
$H^.(U,\CS)$ of the locally constant sheaf $\CS$ of horizontal sections
of $\nabla$.

\subsection{}
\label{irreds} Consider irreducibles $L_i=L(\Lambda_i)$; we denote by
$v_i\in L_i$ the highest vector; set $L=L_1\otimes\ldots\otimes L_n$.
In this Subsection we introduce, following ~\cite{sv}, a certain map
$$
\omega: L_{\Lambda'}\lra \Omega^N
$$

\subsubsection{}
\label{R} The subspace $L_{\Lambda'}$ is generated by all monomials
of the form
$$
x=f_{\vJ_1}v_1\otimes\ldots\otimes f_{\vJ_n}v_n
$$
where $(\vJ_1,\ldots,\vJ_n)$ runs through $Seq^{(n)}(\CJ;\bk)$.

Given such a monomial, we can consider the rational function
$B_{\vJ_1,\ldots,\vJ_n;\pi}(\bt;\bz)$, as in ~(\ref{BBZ}). Set
\begin{equation}
\label{omega}
\omega(x)=B_{\vJ_1,\ldots,\vJ_n;\pi}(\bt;\bz)\cdot l(\bt)\cdot d\bt\in\Omega^N
\end{equation}
where $d\bt:=dt_1\wedge\ldots\wedge dt_N$.

\subsubsection{}
\label{map}
\begin{thm}{} (i) The formula ~(\ref{omega}) correctly defines the map
$$
\omega: L_{\Lambda'}\lra\Omega^N
$$
(ii) We have $\omega((\fn_-L)_{\Lambda'})\subset d\Omega^{N-1}$.
Thus, $\omega$ induces the map
$$
\bomega: L_{\fn_-,\Lambda'}\lra H^N(U,\CS)
$$
\end{thm}

{\bf Proof.} It is one of the main results of ~\cite{sv}, Part II. Cf.
{\em loc. cit}, Cor. 6.13. The key result here is Lemma ~\ref{key}. $\Box$

\subsection{}
\label{statement} Set $\Lambda_{n+1}=\barLam '$. Suppose that
$\Lambda_{n+1}\in C$.
Set $m_0=k-(\Lambda_{n+1},\theta)+1$, $\Lambda''=\Lambda'+m_0\theta$. Consider
the operator
$$
(\bz\cdot f_{\theta})^{m_0}: L_{\Lambda''}\lra L_{\Lambda'}
$$
as in ~\ref{blocks}.

\subsubsection{}
\label{main}
\begin{thm}{} We have $\omega(\Ima((\bz\cdot f_{\theta})^{m_0}))\subset
d\Omega^{N-1}$.

Consequently, if all $\Lambda_i\in C$, $\omega$ induces the mapping
$$
\bomega: W(z_1,\ldots,z_{n+1})\lra H^N(U,\CS)^{\Sigma,-}.
$$
\end{thm}

The rest of this work will be devoted to the proof of the first statement.
Note that it is non-trivial only if $\Lambda''\leq \Lambda$. The last statement
follows from Thm.~\ref{explicit}. $\Box$

\subsection{Resonances} Keep the notations of ~\ref{statement}.

Let us call {\em an edge} any non-empty intersection $L$ of hyperplanes
$H\in\bCC$. Set
$$
a(L)=\sum_{H\in\bCC |H\supset L} a(H)
$$
For instance, consider the point
$L_{\infty}=\{ (\infty,\ldots,\infty)\}\subset X$.

\subsubsection{}
\begin{lem}{} We have
$$
a(L_{\infty})=-\sum_{H\in\CC} a(H)
$$
\end{lem}

{\bf Proof.} Easy. $\Box$

\subsubsection{}
\label{reson}
\begin{lem}{} Suppose that $\Lambda''=\Lambda$. Then
$$
\sum_{H\in\CC} a(H)=-m_0
$$
\end{lem}

{\bf Proof.} From our assumption it follows that $\Lambda'=\Lambda-m_0\theta$.
Recall the notations from ~\ref{notations}. Note that
$m_0=k-(\Lambda',\theta)+1$ since $\Lambda'=\barLam_{n+1}$.

We have $(\Lambda',\theta)=(\Lambda-m_0\theta,\theta)=(\Lambda,\theta)-2m_0$,
so
$$
m_0=k-(\Lambda',\theta)+1=k-(\Lambda,\theta)+2m_0+1,
$$
hence
\begin{equation}
\label{scalar}
(\Lambda,\theta)=k+m_0+1
\end{equation}

On the other hand, one easily sees that
$$
\sum_{1\leq i<j\leq N}a(i,j)=\frac{1}{2\kappa}((m_0\theta,m_0\theta)-
m_0\sum_ia_i(\alpha_i,\alpha_i))=
\frac{1}{\kappa}(m_0^2-m_0(g-1))
$$
since
$$
(\alpha_i,\alpha_i)=<\nu^{-1}(\alpha_i),\alpha_i>=
<a_i^{\vee}a_i^{-1}h_i,\alpha_i>=2a_i^{\vee}a_i^{-1}
$$
It follows that
$$
\sum_{H\in C} a(H)=-\frac{m_0}{\kappa}(\Lambda,\theta)+
\sum_{1\leq i<j\leq N} a(i,j)=-\frac{m_0}{\kappa}(k+g)=-m_0
$$
(cf. ~(\ref{scalar})), and we are done. $\Box$

Now suppose that $\Lambda''\leq\Lambda$. Denote $\beta:=\Lambda-\Lambda''
=\sum_{i=1}^rq_i\alpha_i,\ M:=|\beta|$.

Let us fix maps
$$
p: [M]\lra [r]
$$
such that $\card(p^{-1}(i))=q_i$ for all $i$, and
$$
p': [M+1,N]\lra [r]
$$
such that $\card((p')^{-1}(i))=m_0a_i$ for all $i$ (recall that $\theta=
\sum a_i\alpha_i$). Let us define the map $\pi$, ~(\ref{pi}), as
$$ \pi(j)=\left\{ \begin{array}{ll}
                     p(j) & \mbox{if}\ 1\leq j\leq M \\
                     p'(j) & \mbox{if}\ j>M.
                \end{array}
        \right. $$

The following lemma generalizes ~\ref{reson}.

\subsubsection{}
\label{general}
\begin{lem}{} The sum of $a(H)$ over all $H=H_{ij}\in \CC$ such that
$i$ or $j$ is $>M$ or $H=H_{i;s}\in\CC$ such that $i>M$,
is equal to $-m_0$.
\end{lem}

{\bf Proof.} Computation similar to the one in the proof of ~\ref{reson},
shows that the sum in question is equal to
\begin{equation}
\label{sum}
\frac{m_0}{\kappa}(-(\Lambda,\theta)+m_0-g+1+(\beta,\theta)).
\end{equation}
On the other hand, our assumptions imply that
$$
(\Lambda',\theta)=(\Lambda-\beta-m_0\theta,\theta)=
(\Lambda,\theta)-(\beta,\theta)+2m_0,
$$
hence
$$
m_0=k-(\Lambda',\theta)+1=k-(\Lambda,\theta)+(\beta,\theta)+2m_0+1,
$$
so
$$
m_0=-k+(\Lambda,\theta)-(\beta,\theta)+1
$$
Substituting this into ~(\ref{sum}), we get our claim. $\Box$


\section{Resonance identity}

\subsection{}

\subsubsection{}
\label{setup} We fix $\Lambda_1,\ldots,\Lambda_n\in\fh^*$; we set
$\Lambda=\sum_{i=1}^n\Lambda_i$. We fix non-negative integers $k_1,\ldots,
k_r$, set $\alpha=\sum_{i=1}^rk_i\alpha_i,\ \Lambda'=\Lambda-\alpha$.

We fix a positive integer $m$, and set $\Lambda''=\Lambda'+m\theta$.
We suppose that $\Lambda'\leq\Lambda''\leq\Lambda$. We set
$$
\beta=\Lambda -\Lambda''=\sum_{i=1}^rq_i\alpha_i;\ M=|\beta|
$$
We fix a map
\begin{equation}
\label{p}
\pi_1: [M]\lra [r]
\end{equation}
such that $\card (\pi_1^{-1}(i))=q_i$ for all $i$.

Recall that $\theta=\sum_{i=1}^ra_i\alpha_i$, and all $a_i>0$. We set
$A=|\theta|$. We fix a map
\begin{equation}
\label{prime}
\pi_2: [A]\lra [r]
\end{equation}
such that $\card(\pi_2^{-1}(i))=a_i$ for all $i$.

\subsubsection{} Let us introduce the following sets of independent
variables: $\bu=\{ u_i\}_{1\leq i\leq M};\
\bv (i)=\{ v_j(i)\}_{1\leq j\leq A},\ 1\leq i\leq m,\ \bv =\cup_i\bv (i)$.
Let us assign to every variable $x$ a simple root $\alpha(x)$ as follows.
We set:
$$
\alpha(u_i)=\alpha_{\pi_1(i)};\ \alpha(v_j(i))=\alpha_{\pi_2(j)}
$$
Let us fix distinct complex numbers $z_1,\ldots,z_n$. Let us define
the complex numbers $a(x,y)$, or $a(x,z_s)$ where $x,y$ are any two of
our variables; we will call these numbers {\em exponents}. Namely, set
$$
a(x,y)=(\alpha(x),\alpha(y))/\kappa;\ a(x,z_s)=-(\alpha(x),\Lambda_s)/\kappa
$$
We set
$$
e_{ij}=a(v_i(p),v_j(p))
$$
for any $p\in [m]$ (these numbers do not depend on $p$).

Let us consider the following symbolic expressions.
\begin{equation}
\label{l-lambda''}
l_{\Lambda''}(\bu)=\prod_{i,s}(u_i-z_s)^{a(u_i,z_s)}
\prod_{i,j: i>j}(u_i-u_j)^{a(u_i,u_j)};
\end{equation}
\begin{equation}
\label{l-theta}
l_{\theta}(\bv (p))=\prod_{i\in [A],s\in [n]} (v_i(p)-z_s)^{a(v_i(p),z_s)}
\prod_{i,j\in [A]: i>j}(v_i(p)-v_{j}(p))^{e_{ij}};
\end{equation}
\begin{equation}
\label{l-lambda}
l_{\Lambda'}(\bu,\bv )=l_{\Lambda'}(\bu;\bv (1),\ldots,\bv (m))=
l_{\Lambda''}(\bu)\cdot l'(\bu,\bv )
\end{equation}
where
\begin{equation}
\label{l'}
l'(\bu,\bv )=\prod_p l_{\theta}(\bv(p))\cdot
\prod_{i,j,p}(v_i(p)-u_j)^{a(v_i(p),u_j)}\cdot
\prod_{i,j;\ p,q:\ p>q}(v_i(p)-v_j(q))^{e_{ij}}
\end{equation}

Let us define the following numbers:
\begin{equation}
\label{c}
C(m)=m+\sum\mbox{(all exponents involved in $l'$)};
\end{equation}
\begin{equation}
\label{abcd}
a=\sum_{i,s}a(v_i(p),z_s);\
b=\sum_{i,j:\ i>j}e_{ij};\ c=\sum_{i,j}e_{ij};\
d=\sum_{i,j}a(v_i(p),u_j),
\end{equation}
in the defintion of $a$ and $d$ a number $p\in [m]$ is fixed;
the value does not depend on it.

\subsubsection{}
\begin{lem}{} $C(m)= m+ma+mb+\frac{m(m-1)}{2}c+d$.
\end{lem}

{\bf Proof.} Easy computation. $\Box$

\subsection{} Let $t_1,\ldots,t_n$ be independent variables. Let us define a
differential $(n-1)$-form
\begin{equation}
\label{nu}
\nu(t)=\nu(t_1,\ldots,t_n)=\sum_{i=1}^n(-1)^{i-1}t_idt_1\wedge\ldots\wedge
\hat{dt_i}\wedge\ldots\wedge dt_n
\end{equation}
For any function $f(t)=f(t_1,\ldots,t_n)$ we have
\begin{equation}
\label{homog}
d(f(t)\nu(t))=(nf(t)+\sum_{i=1}^nt\frac{\dpar f}{\dpar t_i})dt
\end{equation}
where $dt:=dt_1\wedge\ldots\wedge dt_n$.

Let us consider a formal expresstion
\begin{equation}
\label{example}
l(t)=l(t_1,\ldots,t_n)=\prod_{i,s}(t_i-z_s)^{p_{is}}\prod_{i>j}
(t_i-t_j)^{q_{ij}}
\end{equation}
Differentiating formally, we get
\begin{equation}
\label{elnu}
d(l(t)\nu(t))=(n+\sum_{i>j}q_{ij}+\sum_{i,s}\frac{p_{is}t_i}{t_i-z_s})l(t)dt
\end{equation}

\subsection{}
It is known that all root spaces of our Lie algebra $\fg$ are one-dimensional.
It follows that a root vector $f_{\theta}\in\fg_{-\theta}$ may be chosen
in the form
\begin{equation}
\label{simple}
f_{\theta}=c_{\theta}\cdot [f_{\vI(\theta)}]
\end{equation}
for some $\vI(\theta)=(i_1,\ldots,i_A)\in [r]^A$ and a non-zero
$c_{\theta}\in \Bbb C$. We fix such a representation once for all.

We also fix an unfolding of $\vI (\theta)$ with respect to $\pi_2$:
$$
\vJ(\theta)=(j_1,\ldots,j_A)\in [A]^A
$$

\subsection{} Let us return to the setup of ~\ref{setup}. Let
$\bt =\{ t_1,\ldots,t_N\}$ denote the union
\begin{equation}
\label{union}
\bt=\bu\cup\bv(1)\cup\ldots\cup\bv(m)
\end{equation}
and
\begin{equation}
\label{big-u}
\btt=\bt\cup\{ z_1,\ldots, z_n\}
\end{equation}

By definition, $N:=\card(\bt)=M+mA$. We order the variables $t_1,\ldots,t_N$
by the natural left-to-right order following from ~(\ref{union}). So,
$$
 (t_1,\ldots, t_M)=(u_1,\ldots,u_M),\ (t_{M+1},\ldots,t_{M+A})=(v_1(1),
\ldots,v_A(1)),
$$
etc.

The maps ~(\ref{p}) and ~(\ref{prime}) induce the surjective map
\begin{equation}
\label{b}
\pi: [N]\lra [r]
\end{equation}
with $\card(\pi^{-1}(i))=q_i+ma_i$ for all $i$. We set $\Sigma=\Sigma_{\pi}$
as in ~\ref{unfold}.

To each variable $t_i\in \bt$ we have assigned a simple root
$\alpha(t_i)=\alpha_{\pi(i)}$, and to a point $z_s$ we assign the
weight $-\Lambda_s$.  We introduce notations
$$
a(i,j):=a(t_i,t_j);\ a(i,s):=a(t_i,z_s).
$$

\subsection{} Note that we can rewrite the expression
{}~(\ref{l-lambda}) in the following form:
\begin{eqnarray}
\label{l(u,v)}
l_{\Lambda'}(\bu,\bv)=l_{\Lambda''}(\bu)\cdot
l_{\theta}(\bv(1))\prod (\bv(1),\bu)\cdot
l_{\theta}(\bv(2))\prod (\bv(2),\bv(1))\prod (\bv(2),\bu)\cdot\ldots\cdot \\
l_{\theta}(\bv(m))\prod (\bv(m),\bv(m-1))\cdot\ldots\cdot\prod(\bv(m),\bv(1))
\prod(\bv(m),\bu) \nonumber
\end{eqnarray}
where
$$
\prod(\bv(i),\bu):=\prod_{p,j}(v_p(i)-u_j)^{a(v_p(i),u_j)}
$$
and
$$
\prod(\bv(i),\bv(j)):=\prod_{p,q}(v_p(i)-v_q(j))^{a(v_p(i),v_q(j))}.
$$

We will denote $l_{\Lambda'}(\bu,\bv)$ simply by $l(\bu,\bv)$, or $l(\bt)$.
We will consider the complex
$$
\Omega_{alg}^.:\ 0\lra\Omega^0_{alg}\lra\ldots\lra\Omega^N_{alg}\lra 0
$$
where $\Omega_{alg}^i$ is the vector space consisting of expressions
$$
\phi(\bt)l(\bt)dt_{p_1}\wedge\ldots\wedge dt_{p_i},
$$
$\phi(\bt)$ being an algebraic rational function of $t_1,\ldots, t_N$.
The differential is defined in the same way as for $\Omega^.$,
{}~(\ref{Omega}). The complex $\Omega^._{alg}$ is naturally
a subcomplex of $\Omega^.$; it inherits the action of the symmetric
group $\Sigma=\Sigma_{\pi}$.

We denote by $\CA: \Omega^._{alg}\lra\Omega^._{alg}$ the operator of skew
symmetrization:
$$
\CA(\omega)=\sum_{\sigma\in\Sigma}(-1)^{|\sigma|}\sigma(\omega)
$$
where $|\sigma|$ denotes the parity of a permutation.

\subsection{} Let us fix $n$ disjoint sequences
$\vI=(\vI_1,\ldots,\vI_n)$ such that $I_1\cup\ldots\cup I_n=[M]$. Set
$\vJ_s=\pi_1(\vI_s)$. We have the corresponding monomial
\begin{equation}
\label{x}
x=x_{\vJ_1,\ldots,\vJ_n}=f_{\vJ_1}v_1\otimes\ldots\otimes f_{\vJ_n}v_n\in
L_{\Lambda''}
\end{equation}

Let us define a differential form
\begin{equation}
\label{derivat}
\tOmega_m=B_{\vI_1,\ldots,\vI_n}(\bu;\bz)\cdot l_{\Lambda''}(\bu)
\cdot d\bu\wedge\tOmega(1)\wedge\ldots
\wedge\tOmega(m)\in\Omega^N_{alg}
\end{equation}
where $\bu:=u_1\wedge\ldots\wedge u_M$, $B_{\vI_1,\ldots,\vI_n}(\bu;\bz)$ is
as in ~(\ref{BZ}), and
\begin{equation}
\label{factor}
\tOmega(i)=c_{\theta}d\{ l_{\theta}(\bv(i))\prod (\bv(i),\bv(i-1))
\cdot\ldots\cdot
\prod (\bv(i),\bv(1))\cdot\prod(\bv(i),\bu))\cdot
A_{\vJ(\theta)}(\bv(i))\cdot\nu(\bv(i))\}.
\end{equation}
Set
\begin{equation}
\label{main-f}
\Omega_m=\CA (\tOmega_m)
\end{equation}

\subsection{} Let us consider a free Lie algebra $\tfn$ with
$N$ generators $\tf_i,\ i\in [M]$ and $^p\tf_a,\ p\in [m], a\in [A]$.

We have elements
$$
\tx:=\tf_{\vI_1}\otimes\ldots\otimes\tf_{\vI_n}\in U(\tfn)^{\otimes n}
$$
and
$$
^p\tf_{\theta}:=c_{\theta}[^p\tf_{\vI(\theta)}]\in\tfn
$$

The construction of Section~\ref{trees} gives us a map
$$
\tB: U(\fn)^{\otimes n}_I\lra R_I(\bz)
$$
for each $I\subset [N]$. Note that
$$
\tB(\tx)=B_{\vI_1,\ldots,\vI_n}(x)
$$

Recall the notation ~(\ref{local}). Pick an $i$-tuple of mutually
distinct numbers $(p_1,\ldots,p_i)\in [m]^i$. By Lemma~\ref{A-B} we have
$$
\tB\{ (\sum_{s=1}^nz_s\cdot^{p_1}\tf_{\theta}^{(s)})\cdot\ldots\cdot
    (\sum_{s=1}^nz_s\cdot^{p_i}\tf_{\theta}^{(s)})\cdot \tx\}=
    W^{(p_1,\ldots,p_i)}(\bu,\bv)\cdot \tB (\tx)\cdot
    A_{\vJ(\theta)}(\bv(p_1))\cdot\ldots\cdot A_{\vJ(\theta)}(\bv(p_i))
$$
for a certain rational function $W^{p_1,\ldots,p_i}(\bu,\bv)$.

Let us describe this function more explicitely.

Let us consider all rational functions of the form
\begin{equation}
\label{adm}
X_{s_1,\ldots,s_i;p_1,\ldots,p_i}(q_1,\ldots,q_i;\tit_1,\ldots,\tit_i)=
z_{s_1}\cdot\ldots\cdot z_{s_i}
\frac{a(v_{q_1}(p_1),\tit_1)}{v_{q_1}(p_1)-\tit_1}\cdot\ldots\cdot
\frac{a(v_{q_i}(p_i),\tit_i)}{v_{q_i}(p_i)-\tit_i}
\end{equation}
where $s_1,\ldots, s_i\in [n],\ q_1,\ldots, q_i\in [A]$ and
$$
\tit_j\in\{ z_{s_j}\}\cup\bu_{I_{s_j}}\bigcup_{j'\in [j-1]|s_{j'}=
s_j}\bv(p_{j'}),\ j=1,\ldots, i
$$
(the last union may be empty). Let us call such
functions {\em admissible terms}.

\subsubsection{} It follows from Lemma~\ref{A-B} that
$W^{p_1,\ldots,p_i}(\bu,\bv)$ is equal to the sum of all admissible
terms with fixed $(p_1,\ldots,p_i)$.

\subsubsection{}
\begin{defn}{} For any $i\in [m]$, the rational function $W_i=W_i(\bu,\bv)$
is defined as a sum
$$
W_i(\bu,\bv)=\sum W^{p_1,\ldots,p_i}(\bu,\bv)
$$
over all $i$-tuples of pairwise distinct numbers $(p_1,\ldots,p_i)\in [m]^i$.
We set $W_0:=1$.
\end{defn}

\subsection{}
\label{main-expl}
\begin{thm}{} We have an equality
\begin{eqnarray}
\label{omega-m}
\Omega_m=c_{\theta}^m l(\bu,\bv)d\bu d\bv\cdot\Sym\{
B_{\vI_1,\ldots,\vI_n}(\bu;\bz)A_{\vJ(\theta)}(\bv(1))\cdot\ldots\cdot
A_{\vJ(\theta)}(\bv(m))\cdot \\
\sum_{i=0}^m (\prod_{j=i+1}^mC(j))W_i(\bu,\bv)\} \nonumber
\end{eqnarray}
\end{thm}

Here
$$
C(j):=j+ja+jb+\frac{j(j-1)}{2}c+d
$$
where $a,b,c,d\in \Bbb C$ are defined in ~(\ref{abcd}).

Equality ~(\ref{omega-m}) will be called  {\bf Resonance identity}.

\subsection{} Let us deduce Theorem~\ref{main} from ~\ref{main-expl}.
{}From the definition ~(\ref{derivat}) follows immediatedly that the
form $\tOmega_m$, and hence $\Omega_m$, is exact. Let us rewrite
{}~(\ref{omega-m}) in the form
\begin{equation}
\label{modified}
\Omega_m=\sum_{i=0}^m(\prod_{j=i+1}^mC(j))\cdot\CW_i(\bu,\bv)
\end{equation}
where
\begin{equation}
\label{cw}
\CW_i(\bu,\bv)=c_{\theta}^m l(\bu,\bv)d\bu d\bv\cdot\Sym\{
B_{\vI_1,\ldots,\vI_n}(\bu;\bz)A_{\vJ(\theta)}(\bv(1))\cdot\ldots\cdot
A_{\vJ(\theta)}(\bv(m))\cdot W_i(\bu,\bv)\}
\end{equation}
Now set $m=m_0$. By Lemma~\ref{general}, $C(m_0)=0$. Thus
$$
\Omega_{m_0}=\CW_{m_0}
$$
On the other hand, it follows from Lemma~\ref{A-B} that
$$
\CW_{m_0}=\omega((\bz\cdot f_{\theta})^{m_0}x)
$$
where $x$ is as in ~(\ref{x}). This proves that the map $\omega$
takes the image of the operator $(\bz\cdot f_{\theta})^{m_0}$ to the subspace
of exact forms, thus proving Theorem\ref{main}. $\Box$

Theorem~\ref{main-expl} will be proved in the next Section.


\section{Proof of Resonance identity}

We keep all the notations of the previous Section.

\subsection{} For each $i\in [m]$, let us consider the operators
"partial differentials"
$$
d^{(i)}:=\sum_{q=1}^A\frac{\dpar}{\dpar v_q(i)}dv_q(i)
$$
acting on our functions or forms.
Note that in the expression ~(\ref{derivat}) we can replace all
forms $\tOmega(i)$ by the forms
\begin{eqnarray}
\label{new}
\tOmega'(i)=c_{\theta}\cdot d^{(i)}\{ l_{\theta}(\bv(i))\cdot
\prod (\bv(i),\bv(i-1))\cdot\ldots\cdot
\prod (\bv(i),\bv(1))\cdot\prod(\bv(i),\bu))\cdot\\
\cdot A_{\vJ(\theta)}(\bv(i))\cdot\nu(\bv(i))\}  \nonumber
\end{eqnarray}
In fact, in the product of the first factor of $\tOmega_m$ and the
first $i-1$ forms $\tOmega(j)$, we have already differentiated the variables
$\bu$ and $v_q(j)$ with $j<i$.

We can apply ~(\ref{elnu}) to ~(\ref{new}), and get
\begin{eqnarray}
\label{better}
\tOmega'(i)=c_{\theta}l_{\theta}(\bv(i))\cdot\prod (\bv(i),\bv(i-1))
\cdot\ldots\cdot
\prod (\bv(i),\bv(1))\cdot\prod(\bv(i),\bu))\cdot \\
\cdot A_{\vJ(\theta)}(\bv(i))\cdot
(1+b+\sum_{q\in [A]; s\in [n]}\frac{a(v_q(i),z_s)v_q(i)}{v_q(i)-z_s}+
\sum_{q\in [A];t\in \bu\cup\bv(1)\cup\ldots\cup\bv(i-1)}
\frac{a(v_q(i),t)v_q(i)}{v_q(i)-t})  \nonumber
\end{eqnarray}
In fact, the expression in brackets in ~(\ref{new}) is the sum
of expressions of the form ~(\ref{example}) with $v_q(i)$ playing the role
of $t$'s in ~(\ref{example}), and $t$'s playing the role of $z$'s.

Thus, we have
\begin{eqnarray}
\label{new-deriv}
\Omega_m=c_{\theta}^m l(\bu,\bv)d\bu d\bv\Sym\{ B_{\vI_1,\ldots,\vI_n}(\bu;\bz)
A_{\vJ(\theta)}(\bv(1))\cdot\ldots\cdot
A_{\vJ(\theta)}(\bv(m))\cdot \\
T(1)\cdot\ldots\cdot T(m)\} \nonumber
\end{eqnarray}
where
\begin{equation}
\label{t}
T(i)=1+b+\sum_{q\in [A]; s\in [n]}\frac{a(v_q(i),z_s)v_q(i)}{v_q(i)-z_s}+
\sum_{q\in [A];t\in \bu\cup\bv(1)\cup\ldots\cup\bv(i-1)}
\frac{a(v_q(i),t)v_q(i)}{v_q(i)-t}
\end{equation}

\subsection{} Let us consider the function $T(m)$. Using the identity
\begin{equation}
\label{gelfand}
\frac{X}{X-Y}=1+\frac{Y}{X-Y}
\end{equation}
many times, one sees that $T(m)$ may be rewritten as
\begin{eqnarray}
\label{T}
T(m)=1+a+b+(m-1)c+d+\sum_{s\in [n],q\in [A]}\frac{z_sa(v_q(m),z_s)}
{v_q(m)-z_s}+ \\
+\sum_{q\in [A]; t\in \bu\cup\bv(1)\cup\ldots\cup\bv(m-1)}
\frac{t\cdot a(v_q(m),t)}{v_q(m)-t} \nonumber
\end{eqnarray}

Let us denote
$$
\tC (m)=1+a+b+(m-1)c+d
$$
Note that $\tC(1)=C(1)$.

Let us consider the expression $\frac{u_j}{v_q(m)-u_j}$ for some
$j\in I_s\subset [M]$, and replace it by
$$
\frac{u_j-z_s}{v_q(m)-u_j}+\frac{z_s}{v_q(m)-u_j}
$$

The next Lemma is our main technical statement. In its proof we use in an
essential way that $\theta$ is the highest root.

\subsection{}
\label{highest}
{\bf Highest root lemma.} {\em We have
$$
\Sym \{ B_{\vI_s}(\bu_{I_s};z_s)\cdot A_{\vJ(\theta)}(\bv(m))\cdot
\sum_{j\in I_s;\ q\in [A]}\frac{u_j-z_s}{v_q(m)-u_j}\}=0
$$}

{\bf Proof.} Suppose $\vI_s=(i_1,\ldots,i_P)$. According to Key Lemma~\ref{key}
we have
$$
B_{\vI_s}(\bu_{I_s};z_s)=\sum_{\sigma\in\Sigma_P}
S(\vI_s,\sigma\vI_s)\cdot E_{\sigma\vI_s}(\bu_{I_s};z_s)
$$
where $\sigma\vI_s:=(i_{\sigma(1)},\ldots,i_{\sigma(P)})$.
Let us consider one product
\begin{equation}
\label{term}
E_{\sigma\vI_s}(\bu_{I_s};z_s)\cdot\frac{u_j-z_s}{v_q(m)-u_j}
\end{equation}
for some $j\in I_s$, say $j=i_{\sigma(p)}$ for some $p\in [P]$, and
rewrite $u_j-z_s$ as a sum
$$
(u_{i_{\sigma(p)}}-u_{i_{\sigma(p-1)}})+\ldots +(u_{i_{\sigma(1)}}-z_s)
$$
The term ~(\ref{term}) becomes a sum of terms corresponding -
in diagrammatic notations of Section~\ref{trees} - to graphs


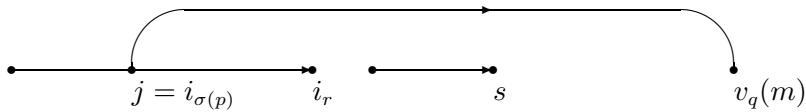
\begin{picture}(15,3)

\put(2,1.5){\circle*{0.15}}
\put(2,1.5){\vector(1,0){5}}
\put(4,1.5){\circle*{0.15}}
\put(4,1){$j=i_{\sigma(p)}$}
\put(7,1.5){\circle*{0.15}}
\put(7,1){$i_r$}

\put(8,1.5){\circle*{0.15}}
\put(8,1.5){\vector(1,0){2}}
\put(10,1.5){\circle*{0.15}}
\put(10,1){$s$}

\put(9,1.5){\oval(10,2)[t]}
\put(9,2.5){\vector(1,0){1}}

\put(14,1.5){\circle*{0.15}}
\put(14,1){$v_q(m)$}

\end{picture}

Now let us fix a decomposition $I_s=I'\cup I''$ with $I'\cap I''=\emptyset$
and $I''\neq\emptyset$, and a sequence $\vI'\in Seq(I')$. Let
$\vI''\in Seq(I'')$ be the total order induced from $\vI_s$. Consider the sum
of terms ~(\ref{term}) (with their coefficients) which produce
the above graphs with the right interval from $i_r$ to $s$ equal to $\vI'$,
the left interval varying.

One sees from \ref{key} that the factors corresponding to the left
interval give a multiple of $A_{\vI''}(\bu_{I''})$. After multiplication
by $A_{\vJ(\theta)}(\bv(m))$ and summing up over all connections between
the left interval and the group $\bv(m)$ and symmetrisation, we get
a multiple of
$$
A([f_{\vI''},f_{\theta}]),
$$
by Lemma~\ref{A-A}. But $[f_{\vI''},f_{\theta}]$ is zero since $\theta$
is a highest root. $\Box$

\subsection{}
\begin{cor}{} Resonance identity is valid for $m=1$.
\end{cor}

{\bf Proof.} In fact, making the above substitution in the expression
for $T(1)$, and taking into account the Highest root lemma, we are
left (after the symmetrisation) with the expression $C(1)+W_1(\bu,\bv(1))$
which gives the Resonance identity for $m=1$. $\Box$

\subsection{} We will proceed with the proof of Resonance identity
by induction on $m$. Suppose we know it for $m-1$. Let us set
$$
D_i^{(m)}=\prod_{j=i+1}^mC(j)
$$
We also denote by $W_i^{(m)}=W_i^{(m)}(\bu,\bv(1),\ldots,\bv(m))$ the function
denoted earlier $W_i(\bu,\bv)$.

By ~(\ref{new-deriv}), we have to show that
\begin{eqnarray}
\Sym\{ B_{\vI_1,\ldots,\vI_n}(\bu;\bz)A_{\vJ(\theta)}(\bv(1))
\cdot\ldots\cdot A_{\vJ(\theta)}(\bv(m))\cdot
T(1)\cdot\ldots\cdot T(m)\}= \nonumber \\
\Sym\{ B_{\vI_1,\ldots,\vI_n}(\bu;\bz)A_{\vJ(\theta)}(\bv(1))
\cdot\ldots\cdot A_{\vJ(\theta)}(\bv(m))\cdot
\sum_{i=0}^mD_i^{(m)}W_i^{(m)}\} \nonumber
\end{eqnarray}
By induction hypothesis, one is reduced to proving that
\begin{eqnarray}
\label{reduce}
\Sym\{ B_{\vI_1,\ldots,\vI_n}(\bu;\bz)A_{\vJ(\theta)}(\bv(1))
\cdot\ldots\cdot A_{\vJ(\theta)}(\bv(m))\cdot
(\sum_{i=0}^{m-1}D_i^{(m-1)}W_i^{(m-1)})\cdot T(m)\}=  \\
\Sym\{ B_{\vI_1,\ldots,\vI_n}(\bu;\bz)A_{\vJ(\theta)}(\bv(1))
\cdot\ldots\cdot A_{\vJ(\theta)}(\bv(m))\cdot
\sum_{i=0}^mD_i^{(m)}W_i^{(m)}\} \nonumber
\end{eqnarray}

By definition,
$$
W_i^{(m-1)}=\sum_{p_1,\ldots,p_i}W^{(m-1)p_1\ldots,p_i}
$$
the sum over all $(p_1,\ldots,p_i)\in [m-1]^i$, $p_j$ mutually distinct.
Let us pick such $(p_1,\ldots,p_i)$. Consider a product of an admissible
term from $W^{(m-1)p_1\ldots,p_i}$ (see ~(\ref{adm})) and a summand from
$T(m)$ (see ~(\ref{T})):
\begin{equation}
\label{product}
X_{s_1,\ldots,s_i;p_1,\ldots,p_i}(q_1,\ldots,q_i;\tit_1,\ldots,\tit_i)\cdot
\frac{a(v_q(m),v_{q'}(j))v_{q'}(j)}{v_q(m)-v_{q'}(j)}
\end{equation}
$(j<m)$. These products occur in the left hand side of ~(\ref{reduce}).

Two cases may occur.

{\em 1st Case.} $j\not\in\{ p_1,\ldots,p_i\}$. Then we can replace the factor
\begin{equation}
\label{factor}
\frac{v_{q'}(j)}{v_q(m)-v_{q'}(j)}
\end{equation}
in the term ~(\ref{product}) in the lhs of ~(\ref{reduce}) by $-\frac{1}{2}$.
In fact, we are doing the symmetrisation which permutes $j$ and $m$, and
we have
$$
\frac{v_{q'}(j)}{v_q(m)-v_{q'}(j)}+\frac{v_{q}(m)}{v_{q'}(j)-v_{q}(m)}=-1.
$$

{\em 2nd Case.} $j=p_r$ for some $r$.

{\bf Claim.} {\em In this case we can replace
{}~(\ref{factor}) by
\begin{equation}
\label{fac}
\frac{z_{s_r}}{v_q(m)-v_{q'}(j)}.
\end{equation}}

In other words, if we substitute
$$
\frac{v_{q'}(j)-z_{s_r}}{v_q(m)-v_{q'}(j)}
$$
into the lhs of ~(\ref{reduce}), we get $0$ after symmetrisation.
This claim is proved by the argument identical to the argument
in the proof of Highest root lemma ~\ref{highest}.

Let us denote for brevity
$$
Y(\bt;\bz)=B_{\vI_1,\ldots,\vI_n}(\bu;\bz)A_{\vJ(\theta)}(\bv(1))
\cdot\ldots\cdot A_{\vJ(\theta)}(\bv(m))
$$
Using Highest root lemma, we can rewrite the lhs of ~(\ref{reduce})
as
\begin{eqnarray}
\label{lreduce}
\Sym\{ Y(\bt;\bz)\cdot \sum_{i=0}^{m-1}D_i^{(m-1)}\sum_{p_1,\ldots,p_i}
W^{(m-1)p_1,\ldots,p_i}
\cdot (\tC(m)+\sum_{s,q}(\frac{a(v_q(m),z_s)z_s}{v_q(m)-z_s}+ \\
\sum_{j\in I_s}\frac{a(v_q(m),u_j)z_s}{v_q(m)-u_j})+
\sum_{j\in [m-1]-\{ p_1,\ldots,p_i\} }
\sum_{q,q'}(-\frac{a(v_q(m),v_{q'}(j))}{2})+
\sum_{r=1}^i\sum_{q,q'}\frac{a(v_q(m),v_{q'}(p_r))z_{s_r}}{v_q(m)-v_{q'}(p_r)})
\} = \nonumber \\
\Sym\{ Y(\bt;\bz)\cdot \sum_{i=0}^{m-1}D_i^{(m-1)}\sum_{p_1,\ldots,p_i}
W^{(m-1)p_1,\ldots,p_i}
\cdot (\tC(m)+\sum_{s,q}(\frac{a(v_q(m),z_s)z_s}{v_q(m)-z_s}+ \nonumber \\
\sum_{j\in I_s}\frac{a(v_q(m),u_j)z_s}{v_q(m)-u_j})
-\frac{m-i-1}{2}c+
\sum_{r=1}^i\sum_{q,q'}\frac{a(v_q(m),v_{q'}(p_r))z_{s_r}}{v_q(m)-v_{q'}(p_r)})
\} = \nonumber
\end{eqnarray}
We have to prove that ~(\ref{lreduce}) is equal to
\begin{equation}
\label{rreduce}
\Sym\{ Y(\bt;\bz)\cdot \sum_{i=0}^mD_i^{(m)}\sum_{p'_1,\ldots,p'_i}
W^{(m)p'_1,\ldots,p'_i}\}
\end{equation}

\subsection{} Let us consider more attentively the nature of symmetrisation.
Let us denote by
$$
\pi^{(m)}:[N]=[M+mA]\lra [r]
$$
the map ~(\ref{b}), and by
$$
\pi^{(i)}:[M+iA]\lra [r]
$$
the analogous map with $m$ replaced by $i\in [m]$. Denote
$$
\Sigma^{(i)}=\Sigma_{\pi^{(i)}}
$$
The symmetrisation in ~(\ref{lreduce}),~(\ref{rreduce}) is done over the group
$\Sigma=\Sigma^{(m)}$.

Note that $\Sigma^{(m)}$ is equal to a disjoint union
$$
\Sigma^{(m)}=\bigcup_{i=1}^m\Sigma^{(m-1)}\cdot (im)
$$
where $(im)\in\Sigma^{(m)}$ denotes the transposition of the whole group
$\bv(i)$ with $\bv(m)$.

More generally, the symmetric group $\Sigma_m$ is naturally embedded
in $\Sigma^{(m)}$ - it acts by permutations of groups $\bv(i)$.
This subgroup evidently fixes $Y(\bt;\bz)$.

Let us denote
$$
Q=\{ (1m),\ldots, (mm)\}\subset \Sigma_m\subset\Sigma^{(m)}
$$
We have
$$
\Sym_{\Sigma^{(m)}}=\Sym_{\Sigma^{(m-1)}}\circ \Sym_Q
$$
Let us pick mutually distinct $(p_1,\ldots,p_i)\in [m-1]^i$, and consider
the partial symmetrisation of the corresponding summand in
{}~(\ref{lreduce}):
\begin{eqnarray}
\label{lra}
\Sym_Q\{ Y(\bt;\bz)\cdot \sum_{i=0}^{m-1}D_i^{(m-1)}
W^{(m-1)p_1,\ldots,p_i}
\cdot (\tC(m)+\sum_{s,q}(\frac{a(v_q(m),z_s)z_s}{v_q(m)-z_s}+ \\
\sum_{j\in I_s}\frac{a(v_q(m),u_j)z_s}{v_q(m)-u_j})
-\frac{m-i-1}{2}c+
 \sum_{r=1}^i\sum_{q,q'}\frac{a(v_q(m),v_{q'}(p_r))z_{s_r}}
{v_q(m)-v_{q'}(p_r)})\} = \sum_{j=1}^m S_j \nonumber
\end{eqnarray}
where
\begin{eqnarray}
S_j=  Y(\bt;\bz)\{ \cdot \sum_{i=0}^{m-1}D_i^{(m-1)}
W^{(m-1)p_1,\ldots,p_i}
\cdot (\tC(m)+\sum_{s,q}(\frac{a(v_q(m),z_s)z_s}{v_q(m)-z_s}+ \\
\sum_{j\in I_s}\frac{a(v_q(m),u_j)z_s}{v_q(m)-u_j})
-\frac{m-i-1}{2}c+
 \sum_{r=1}^i\sum_{q,q'}\frac{a(v_q(m),v_{q'}(p_r))z_{s_r}}
{v_q(m)-v_{q'}(p_r)})\}^{(jm)} \nonumber
\end{eqnarray}
Consider the $j$-th summand $S_j$. Two possibilities
may occur:

(i) $j\in [m]-\{ p_1,\ldots, p_i\}$. In this case
$$
S_j=Y(\bt;\bz)\cdot (D_i^{(m-1)}\cdot (\tC(m)-\frac{m-i-1}{2}\cdot c)\cdot
W^{(m)p_1,\ldots,p_i}+D_i^{(m-1)}\cdot W^{(m)p_1,\ldots,p_i,j})
$$

(ii) $j=p_r$ for some $r$. Then
$$
S_j=Y(\bt;bz)\cdot (D_i^{(m-1)}\cdot (\tC(m)-\frac{m-i-1}{2}\cdot c)\cdot
W^{(m)p_1,\ldots,p_{r-1},m,p_{r+1},\ldots, p_i}+
D_i^{m-1}\cdot W^{(m)p_1,\ldots,p_i,m}),
$$
as one sees from definitions.

Now, if we pick mutually distinct $(p'_1,\ldots,p'_i)\in [m]^i$, we see that
the contribution into $W^{(m)p'_1,\ldots,p'_i}$ from ~(\ref{lra}) comes with
a coefficient
\begin{eqnarray}
\label{coeff}
D_i^{(m-1)}\cdot (\tC(m)-\frac{m-i-1}{2}\cdot c)\cdot (m-i)+ D_{i-1}^{(m-1)}=
\\
D_i^{(m-1)}\cdot ((m-i)\cdot (\tC(m)-\frac{m-i-1}{2}\cdot c)+ C(i))
\nonumber \\
\end{eqnarray}

\subsection{}
\begin{lem}{} We have
$$
C(m)=(m-i)\cdot (\tC(m)-\frac{m-i-1}{2}\cdot c)+ C(i)
$$
\end{lem}

{\bf Proof.} Immediate. $\Box$

It follows that ~(\ref{coeff}) is equal to $D_i^{(m-1)}C(m)=D_i^{(m)}$.
It follows that (\ref{lreduce})=(\ref{rreduce}) which in turn implies
$m$-th Resonance identity. $\Box$



\end{document}